\documentclass[12pt]{article}
\usepackage{amsmath,amssymb,latexsym}

\def\hybrid{\topmargin -20pt  \oddsidemargin 0pt
          \headheight 0pt   \headsep 0pt
          \textwidth 6.25in % A4 paper
          \textheight 9.5in % A4 paper
          \marginparwidth .875in
          \parskip 5pt plus 1pt   \jot = 1.5ex}

\hybrid

\def\x{\times}

\def\o+{\oplus}
\def\ra{\rightarrow}
\def\da{\downarrow}
\def\lra{\longrightarrow}

\def\beqa{\begin{eqnarray}}
\def\eeqa{\end{eqnarray}}
\def\del{\partial}
\def\pa{\partial}
\def\al{\alpha}

\def\si{\sigma}
\def\th{\theta}

\def\G{\Gamma}

\def\cA{{\cal A}}
\def\cB{{\cal B}}

\def\F{{\cal F}}
\def\J{{\cal J}}

\def\S{{\cal S}}
\def\T{{\cal T}}

\def\Tb{\bar{\T}}

\def\Re{{\text{Re}}}
\def\Im{{\text{Im}}}

\def\ov{\overline}
\def\un{\underline}
\def\C{\tilde{C}}

\def\jb{{\bar\jmath}}
\def\p{\partial}

\newcommand{\resetcounter}{\setcounter{equation}{0}}

\begin{document}

\thispagestyle{empty}
\rightline{LMU-ASC 40/06}
\rightline{HD-THEP-06-10}
\rightline{hep-th/0606047}
\vspace{.8cm}
\centerline{\bf \LARGE On the modified KKLT procedure:}
\vspace{.5truecm} 
\centerline{\bf \LARGE a case study for the ${\bf P}_{11169}[18]$ model}

\vspace{1truecm} \centerline{Gottfried Curio$^{a}$ and Vera Spillner$^b$}

\vspace{.7truecm}

\centerline{{\em $^a$Arnold-Sommerfeld-Center for Theoretical
Physics}} \centerline{{\em Department f\"ur Physik,
Ludwig-Maximilians-Universit\"at M\"unchen}} 
\centerline{{\em Theresienstra\ss e 37, 80333 M\"unchen, Germany}}

\vspace{.2truecm}

{\em \centerline{$^b$Institut f\"ur Theoretische Physik, 
Universit\"at Heidelberg}} 
{\em \centerline{Philosophenweg 16, 69120 Heidelberg, Germany }}

\vspace{.2truecm}

%%%%%%%%%%%%%%%%%%%%%%%%%%%%%%%%%%%%%%%%%%%%%%%%%%%%%%%%%

\begin{abstract}
We probe the existence of supersymmetric vacua of the type IIB orientifold 
of the elliptic Calabi-Yau space ${\bf P}_{11169}[18]$ where
generically two complex structure moduli $z_i$, the dilaton $\tau$ and the
two K\"ahler moduli $T_i$ are stabilized by fluxes and gaugino condensates. 
The usual KKLT procedure,
which integrates out the complex structure moduli {\em and the dilaton},
actually has to be modified, such that one keeps the dependence on $\tau$.
We derive explicitely the resulting effective superpotential 
$W_{eff}(\tau)$ for the dilaton for various flux combinations.
As this is actually a non-holomorphic quantity one must properly work 
with the $G$-function. The remaining SUSY equations for $\tau$ and the
$T_i$ can be resolved explicitely.

\end{abstract}

\newpage
\pagenumbering{arabic}

\section{Introduction}

The issue of moduli stabilization in superstring theory gained new interest
when it was observed [\ref{GKP}], [\ref{fluxes}]
that by adding $3$-form fluxes, inducing a superpotential
$W_{flux}$, in the type IIB theory on a Calabi-Yau $X$
the complex structure moduli $z_i$ and the dilaton $\tau$ can be frozen.
In a corresponding orientifold model, reducing $4D$ space-time
supersymmetry to $N=1$, the presence of orientifold planes leads to
cancellation of $D$-brane charges against the effects of $D3/D7$-branes 
and $3$-form fluxes in the tadpole cancellation condition.

KKLT [\ref{KKLT}]  combined this with 
a non-perturbative superpotential $W_{np}$
describing gaugino condensation [\ref{DRSW}]
(or Euclidean $D3$-instantons [\ref{W}]) such that
the remaining K\"ahler moduli $T_i$ could be stabilized also.
Here, for warped $D7$-branes, the non-abelian gauge theory on 
a stack of $D7$-branes must allow for a non-vanishing gaugino condensate; 
for Euclidean $D3$-brane instantons the number of 
fermionic zero-modes on the world-volume is relevant (here the fluxes 
have a subtle impact on the relevant count [\ref{Kallosh}]).

More precisely, the scenario of KKLT amounts 
(working at supergravity level at large $\tau$ and $T_i$)
to a two-step procedure where $z_i$ and $\tau$ 
have been fixed by fluxes first, 
leading just to a constant effective superpotential contribution $W_0$, 
and in a second step,
in the resulting effective theory, the K\"ahler moduli $T_i$ are stabilized 
(this decoupling procedure assumes that the $z_i$ and $\tau$ are much heavier
than the $T_i$ so that they can be integrated out first). The resulting
supersymmetric $AdS$-vacuum was then uplifted to a non-supersymmetric 
$dS$-vacuum by adding anti $D3$-branes (for whose stability the (mass)$^2$
eigenvalues of the fixed scalars should be already positive in the 
$AdS$-vacuum and not just negative and fulfilling the
Breitenlohner-Freedman bound).

In [\ref{Nilles}] stability conditions were derived 
(for a susy $AdS$-vacuum with stabilized moduli)
in a more general case than the decoupling limit of KKLT. There only the $z_i$ 
(if heavy enough) were integrated out, 
leading to an effective superpotential contribution $W_{eff}(\tau)$
%from the flux superpotential
\beqa
\label{Wflux general}
W_{flux}(z_i, \tau)=A(z_i)+\tau B(z_i)\;\; 
\stackrel{D_i W_{flux}=0}{\leadsto}\;\;
W_{eff}(\tau)
\eeqa 
(define for later $\gamma:=\tau_2 W_{eff}''/W_{eff}'$).
There as ansatz for the K\"ahler potential is used
\beqa
\label{K general}
K^{(T)}(T)+K(\tau)=-3\log(T+\bar{T})-\log(\tau - \bar{\tau})
\eeqa
The supersymmetric stationary points fulfill $D_i W=0$.
Their nature is determined by the second derivatives $V''$ 
of the scalar potential of the combined superpotential
$W=W_{eff}(\tau)+Ce^{-aT}$ (here in principle the $C$ was also 
$z_i$-dependent, but is set constant for now): 
this matrix splits into 'real' and 'imaginary' (axionic)
parts $V_{0M}''$ and $V_{0A}''$ and
the stability condition for $V$ from the matrix determinants 
(at large $T$) is $|\gamma|> 1$. 

In [\ref{de Alwis}] it was pointed out that this program should
be properly enhanced to take into account two points. First, the
K\"ahler potential $K(z_i)$ of the complex structure moduli, more precisely
the effective version $K_{eff}(\tau)=K\big(z_i(\tau)\big)$ of it 
arising after integrating out
the $z_i$, has to be added to $K(\tau)=-\ln (\tau - \bar{\tau})$. Secondly,
as the conditions $D_i W_{flux}=0$ for integrating out the $z_i$
are (because of the covariant derivative involving the K\"ahler potential)
actually non-holomorphic, it turns out that the effective superpotential
$W_{eff}(\tau)$ arising from $W_{flux}(\tau, z_i)$ is also actually a 
non-holomorphic quantity. Therefore one should compute the scalar potential
in the form $V=e^{G}\big(G^{i\bar\jmath}G_i G_{\bar\jmath}-3\big)$
where
\beqa 
G=K(\tau)+K_{eff}(\tau)+K(T)+\ln|W_{eff}+W^{(T)}|^2
\eeqa
and the $z_i$ should be integrated out by $G_i=0$ or equivalently 
(for $W\neq 0$) $D_i (W_{flux}+W^{(T)})=0$, leading actually to
$z_i(\tau, T)$ instead of just $z_i(\tau)$. This is described in sect.~5.

Just as in the case of KKLT this program now should be carried out 
for concrete examples. We have choosen the elliptic Calabi-Yau 
$X={\bf P}_{11169}(18)$ and 
its corresponding orientifold model for which the KKLT program 
was carried through [\ref{DDF}], cf. also [\ref{8authors}], [\ref{CoQu}], 
[\ref{BaBe}]. For some examples of choices 
of fluxes we derive here $W_{eff}$ and $K_{eff}$ in completely 
explicit form. For one fully worked out case 
we find that no supersymmetric solutions exist. 
For the procedures cf.~sect.~6.
Here we were just searching for SUSY solutions;
more generally one would search for non-SUSY stationary points of the 
resulting $V$.

In one concrete 
example we take the flux combination (where we specialise to $c=3d$)
\beqa
(e_R^a \, | \, m_R^a)= (0, g , 0 \, | \, 0 , -a , -b)
\;\;\; , \;\;\;
(e_{NS}^a \, | \, m_{NS}^a) = (0 , 0 , 0 \, | \, 0 , c , d)
\eeqa
which gives for $W_{flux}=gF_1+ az_1+bz_2+\tau(cz_1+dz_2)$
the explicit expression
\beqa
W_{flux}(z_i, \tau)=g\big(-\frac{1}{2}(3z_1+z_2)^2 + \frac{3}{2}(3z_1+z_2)
+\frac{17}{4} \big)+(a+\tau c)z_1+(b+\tau d)z_2
\eeqa
The K\"ahler potential of the complex structure moduli $z_i$ is given by
(where $y_k=\mbox{Im}z_k$)
\beqa
K(z_1, z_2) 
=- \ln \Big[ y_1\big(3y_1^2+3y_1y_2+y_2^2\big)+i\xi\Big]
\eeqa
The fourfold we work on following [\ref{DDF}] has a 
base $B_3$, itself ${\bf P^1_z}$ fibered over $B_2={\bf P^2}$; the latter 
is embedded as divisor $r$ and $r_{\infty}$ in $B_3$ at $z=0$ and
$z=\infty$, respectively, and has a $G_2$ singularity along $r$
and an $E_6$ or $E_8$ singularity along $r_{\infty}$.
With the volumes  $\tau_1:=\tau_r=\frac{1}{2}t_1^2 $ 
and $\tau_2:=\tau_{r_{\infty}}=\frac{1}{2}(t_1+6t_2)^2$ 
of the divisors $r$ and $r_{\infty}$ a superpotential 
\beqa
W^{(T)}=C_1 e^{-2\pi a_1 \T_1} + C_2 e^{-2\pi a_2 \T_2}
\eeqa
is generated (with the complexified 4-cycle-volumes
$\T_j=\tau_j+i \th_j$ where $\th_i=-\int_{r_{(\infty)}}C_4$).
For the K\"ahler form $J=t_1 L + t_2 r_{\infty}$ (cf.~ below)
one has the K\"ahler potential (so $\tau_2 > \tau_1$)
\beqa
K(t_i)=-2\ln(\tau_2^{3/2}-\tau_1^{3/2})
\eeqa

In {\em section 2} we consider the $N=2$ theory from which 
the orientifold model is derived. We give the prepotential, 
the K\"ahler potential and the period vector for the complex structure moduli, 
leading to a concrete expression of $W_{flux}$. 
In {\em section 3} we give the $N=1$ orientifold theory 
with its K\"ahler potential. In {\em section 4} we recall 
the KKLT procedure for the model.
In {\em section 5} we show how one works
for $W_{eff}$ non-holomorphic actually with the $G$ function.
In {\em section 6} we derive $W_{eff}(\tau, \T_k)$ for various
flux combinations and give a complete analytical treatment of the remaining 
SUSY conditions for $\tau$ and the $\T_k$.

%\newpage
\section{The $N=2$ parent theory on the Calabi-Yau $X_3$}

\resetcounter

We consider the elliptic Weierstrass fibration $\pi: X\ra B_2$
over $B_2={\bf P^2}$. Some data of such elliptic
Calabi-Yau threefolds (with section $\sigma$) over $B_2$ are 
(by adjunction) [\ref{FMW}]
\beqa
\label{elliptic relations 1}
\si^2&=& -c_1 \si \\
\label{elliptic relations 2}
c_2(Z)&=&12 c_1 \si + c_2 + 11 c_1^2 
\eeqa
One has $h^{1,1}(X)=2$ from the divisors $\si$ (the embedded base 
of the fibration) and $L:=\pi^{-1}\, l$ 
where $l$ is the line in $B_2$. Furthermore
$c_3(X)=-60 c_1^2$ gives $h^{2,1}(X)=272$.

{\em Classical intersection numbers on $Z$}

With $H:=\si+3L$ and $L$ as divisors 
one finds using (\ref{elliptic relations 1})
\beqa
H^3=9 \, , \;\;\;\; H^2L=3 \, , \;\;\;\; HL^2=1 \, , \;\;\;\; L^3=0
\eeqa
Furthermore one gets from (\ref{elliptic relations 2}) that
that
\beqa
c_2(Z)\cdot L = 36\;\;\; , \;\;\;
c_2(Z)\cdot H = 108 + (-12 c_1^2+c_2+11c_1^2)=102
\eeqa

{\em K\"ahler potential}

Let us give the corresponding K\"ahler potential for the 
K\"ahler form $\tilde{\J}:=t_1H + \tilde{t}_2 L$
\beqa
\label{CY Kahler pot H L}
e^{-K}=V=\frac{1}{6}\Big( 9t_1^3+9t_1^2\tilde{t}_2+3t_1\tilde{t}_2^2\Big)
\eeqa

{\em Remark:} This is the K\"ahler potential in the basis
adapted to the prepotential given below. Later, when we consider
the orientifold $N=1$ theory derived from the $N=2$ Calabi-Yau theory,
the relevant divisors to consider will be suitable quotients
related to $\sigma$ (then called $r$) and $L$. For this reason let
us also give here already the K\"ahler potential for the 
K\"ahler form $\J':=t_1\sigma + t^{'}_2 L$, i.e. with intersection numbers 
\beqa
\sigma^3=9 \, , \;\;\;\;  \sigma^2 L &=& -3
\, , \;\;\;\;  \sigma L^2 = 1\, , \;\;\;\;  L^3=0 \\
%\eeqa
%This gives 
%\beqa
\label{CY Kahler pot sigma L}
e^{-K}=V&=&\frac{1}{6}\Big( 9t_1^3-9t_1^2t^{'}_2+3t_1(t^{'}_2)^2\Big)
\eeqa

\newpage

{\em The prepotential and the periods}

Expressing the middle homology and cohomology in 
terms of a symplectic basis, 
i.e.~a basis of 3-cycles $A^a$ and $B_b$ 
with $A^a \cap B_b = -B_b \cap A^a  =  \delta^a_b$ and
$A^a \cap A^b = B_a \cap B_b  = 0$
and a basis of 3-forms $\alpha_a$ and $\beta^b$ 
(where $a,b=0,1, \ldots , h^{2,1}$; let $n=h^{2,1}+1$)
\beqa
\int_{A^b} \alpha_a = - \int_{B_a} \beta^b  =  \delta^b_a \;\; ,
\;\;\;\;\;\;\;\;
\int_{X} \alpha_a \wedge \beta^b = -\int_{X} \beta^b
\wedge
\alpha_a  =  \delta^b_a 
\eeqa
Such a symplectic basis is only defined up to $Sp(2n, {\bf Z})$
transformations, 
as these preserve the symplectic intersection form.
The periods are defined via the holomorphic 3-form 
$\Omega$ 
\beqa
\int_{A_a} \Omega = X^a\, ,\qquad \qquad
   \int_{B^b} \Omega = \F_a
\eeqa
The periods are collected in the period vector, $\Pi =
\left( X^0,\ldots, X^n , F_0, \ldots ,F_n \right)$. 
This is a function of the complex structure moduli
$z^i=X^i/X_0$ ($i=1, \dots , h^{2,1}$)
and inherits the holomorphic freedom 
of $\Omega$ and is defined up to holomorphic rescalings $\Omega \to
f(z_i)\Omega$.

Here the prepotential $\F$ is given by 
(with $\xi'=\frac{\zeta(3)}{2(2\pi i)^3}\chi(X)\approx -1.3 i$ 
and $q_i=e^{2\pi i z_i}$)
\beqa
\F&=&-\frac{C_{ijk}}{3!}\frac{X^iX^jX^k}{X^0}+\frac{A_{ij}}{2}X^iX^j+
c_i X^iX^0 - \xi' (X^0)^2+f(q)(X^0)^2\nonumber\\
&=&(X^0)^2 \F=(X^0)^2
\Bigg( -\frac{C_{ijk}}{3!}z^iz^jz^k+ \frac{1}{2} A_{ij}z^iz^j+
c_i z^i - \xi' +f(q) \Bigg)
\eeqa
where $C_{ijk}=\int J_i\wedge J_j\wedge J_k=D_i\cap D_j\cap D_k$ 
are the classical intersection numbers and 
$c_i=\frac{1}{24}\int c_2(X)J_i$ (the quadratic coeffecients $A_{ij}$
are unphysical and not completely fixed).

In our case this becomes 
(we take later $A_{11}=9/2, A_{12}=3/2, A_{22}=0$ 
with [\ref{CFKM}], [\ref{DDF}])
\beqa
\F=-\frac{1}{6}(9z_1^3+9z_1^2z_2+3z_1z_2^2)
+\frac{9}{4}z_1^2+\frac{3}{2}z_1z_2
+\frac{17}{4}z_1+\frac{3}{2}z_2-\xi'
\eeqa

With this prepotential the period vector becomes
(where $F_a=\pa F / \pa X^a$ and $\F_i = \pa \F / \pa z^i$ for
the prepotential $F(X^a)$ in homogeneouos coordinates
($a=0,1, \dots , h^{2,1}$) and $\F(z^i)$ in inhomogeneouos
coordinates $z^i=X^i/X^0$ ($i=1,\dots , h^{2,1}$), respectively)
\beqa
\label{period vector}
\Pi = \left( \begin{array}{c} X^0 \\ X^i \\ F_0 \\ F_i \end{array} \right) =
X^0\left(\begin{array}{c} 1 \\ z^i \\ 2\F - z^i \F_i \\ \F_i
\end{array}\right)
=X^0\left(\begin{array}{c} 1 \\ z^i \\ 
\frac{C_{ijk}}{3!}z^iz^jz^k +c_i z^i - 2 \xi' + \dots \\
-\frac{C_{ijk}}{2}z^jz^k + A_{ij}z^k + c_i + \dots \end{array}\right)
\eeqa
The K\"ahler potential for the complex
structure moduli space is given by 
\beqa
K(z_i)& = &- \ln \left( i \int
\Omega \wedge \bar{\Omega} \right)=- \ln \left( -i\Pi^\dagger
\cdot \Sigma \cdot \Pi \right) \nonumber\\
&=&-\ln \Big( i |X^0|^2 ((z_i-\bar{z_i})(\F_i + \bar{\F_i}) - 2 (\F -
\bar{\F}))\Big)
\eeqa
with 
$\Sigma = \tiny{
\left( \begin{array}{cc} 
0 & \bf{1}_n \\ -\bf{1}_n & 0 
\end{array} \right)}$.
Further one has in the dilaton sector
$K(\tau) =   -\ln(-i(\tau - \bar{\tau}))$.

\newpage

One calculates explicitly
(with $z_k=x_k+iy_k$; $\xi:=-\xi'$)
\beqa
\label{z_i potential}
K(z_1, z_2) 
=- \ln \Big( y_1\big(3y_1^2+3y_1y_2+y_2^2\big)+i\xi\Big) 
\eeqa
(up to an additive constant).
Compare the K\"ahler potential (\ref{z_i potential}) 
for the complex structure moduli (expressed in the $y_k$)
with the K\"ahler potential (\ref{CY Kahler pot H L}) for the K\"ahler moduli
(expressed in $t_1=\Re \, T_1$, $\tilde{t}_2=\Re \, T_2$ 
and with respect to $H$ and $L$)
\beqa
K(t_1, \tilde{t}_2)=-\ln \Big( \frac{1}{2}\big(
3t_1^3+3t_1^2\tilde{t}_2+t_1\tilde{t}_2^2\big)\Big)
\eeqa
Restricting the consideration of the complex structure moduli
to two of them (as recalled below) one gets the same K\"ahler potential
as for the K\"ahler moduli by mirror symmetry.

Note that $Z$ has the (weighted) projective embedding as hypersurface
${\bf P}_{11169}[18]$
\beqa
\label{18}
x_1^{18}+x_2^{18}+x_3^{18}+x_4^3+x_5^2
-18\psi x_1x_2x_3x_4x_5-3\phi x_1^6x_2^6x_3^6=0
\eeqa
Here actually only $2$ of the full set of $272$ deformations are displayed.
The group $\Gamma={\bf Z_6}\x {\bf Z_{18}}$ is a symmetry of 
the moduli space and fixes the two-parameter subspace in (\ref{18}).
One is lead to this specific Calabi-Yau by the mirror construction
and the periods of $\Gamma$-invariant cycles are its six periods.
One proceeds to find flux-vacua by working in the $\Gamma$-invariant
part of the moduli space (turning on only invariant fluxes); i.e. one
turns on fluxes just on these cycles and has $D_i W=0$ in non-invariant
directions as the resulting superpotential and K\"ahlerpotential are
invariant. Effectively
it is thus consistent to set all other moduli to zero and work only on
the moduli in (\ref{18}) and their associated fluxes.

\subsection{The flux superpotential}

In terms of the periods, the flux-superpotential is\footnote{note that
for two three-forms one has
$\int \phi\wedge \chi = 
\sum_a \int_{A^a}\phi \int_{B_a}\chi - \int_{B_a}\phi \int_{A^a}\chi$}
\beqa
\label{GWV supo}
W_{G} &=& \int_X G_3 \wedge \Omega 
= (2 \pi)^2 \alpha' \Big( (e_R^a - \tau e_{NS}^a)\cdot \Pi_{3+a} 
- (m_R^a - \tau m_{NS}^a)\cdot \Pi_{a} \Big)\nonumber\\
&=& (2 \pi)^2 \alpha' \Big( \sum_{a=0}^k (e_R^a - \tau e_{NS}^a)F_a - 
(m_R^a - \tau m_{NS}^a)z_a\Big)
\eeqa
where the
integral vectors of fluxes along the cycles occur as precoefficients. 
More precisely the fluxes $H_R$ and
$H_{NS}$ are elements of $H^3(Z, {\bf Z})$ quantised as follows
\beqa
e_{R/NS}^a = 
\frac{1}{(2 \pi)^2 \alpha'} \int_{A^a} H_{R/NS} \in {\bf Z}\; , \;\;\;\;
(m_{R/NS})_a = 
\frac{1}{(2 \pi)^2 \alpha'} \int_{B_a} H_{R/NS} \in {\bf Z}
\eeqa
This amounts to a $D3$-charge carried by the fluxes 
\beqa
N_{flux} = L
=\frac{1}{2}(e^a_R,m^a_R) \cdot \Sigma \cdot (e^a_{NS},m^a_{NS})^t
=\frac{1}{2}\sum_{a=0}^k e_R^a\, m_{NS}^a-m_R^a\, e_{NS}^a
\eeqa

\newpage

\section{The $N=1$ theory and the orientifold limit of the fourfold}

\resetcounter

Let $\pi: X_4 \ra B_3$ denote the projection of the elliptic fibration
of a Calabi-Yau fourfold in $F$-theory.
We consider the case that $B_3$ is ${\bf P^1}$-fibered over the base $B_2$. 
In the orientifold limit a $D4$ singularity is realized.

In the type IIB string theory one describes
$B_3$ as quotient by a holomorphic involution 
of a Calabi-Yau threefold $Z$,
branched at the $B_3$-location of the singularities 
of the elliptic fibration of $X_4$ ($D_4$ singularities corresponding to an 
$O7$-plane and four coincident $D7$-branes).
$D$-branes are introduced to cancel the $RR$ tadpoles 
produced by the orientifold plane. 
If $Z$
is given as a hypersurface in an ambient ${\bf P^4}$ this involution
might be given by $z_1\ra - z_1$.

The dilaton is the complex structure modulus
of the elliptic $F$-theory fibre.
\beqa
Z \;\;\;\;& \lra & \;\;\; B_3\nonumber \\
\;\;\;\;  \da \, E & & \;\; \;\; \da \, {\bf P^1} \;\;\; \nonumber \\
B_2 \;\;\; & = & \;\;\; B_2 
\eeqa

$B_3$ inherits the K\"ahler and complex structure moduli
which are even and odd, respectively, under the involution.

{\em Classical intersection numbers on $B_3$}

The ${\bf P^1}$-fibration of $B_3$ over $B_2$ 
can be described as projectivization ${\bf P}(Y)$ of a vector bundle 
$Y={\cal O}\oplus {\cal T}$ with ${\cal T}$ a line bundle over $B_2$.

Let ${\cal O}(1)$ denote a line bundle on the total space which 
restricts on each ${\bf P^1}$ fibre to the corresponding line bundle 
over ${\bf P^1}$ and let $r=c_1({\cal O}(1))$ and $t=c_1({\cal T})$ 
so that the cohomology of $B_3$ is generated over $B_2$ by $r$
with the relation $r(r+t)=0$ (as the two section are disjoint:
one has $r=r_0$ the divisor given by $B_2$ itself inside $B_3$,
at the zero-section of the ${\bf P^1}$-fibration;
the section at infinity is given by $r_{\infty}=r+t$).
Concretely take $B_2={\bf P^2}$ and $t=6 l$. Here we denote the line in
$B_2$ 
by $l$ and its preimage in $B_3$ by $L={\bf F_6}$ 
(one has $Lr^2=(r|_L)^2=(l)^2_L=-6$; furthermore $Lr_{\infty}^2=6$).

One has the following intersection numbers in $B_3$
(the first three relations are obvious)
\beqa
L^3=0\, , \;\;\;\; L^2r=L^2r_{\infty}=1\, , \;\;\;\; Lr_{\infty}^2=6 \, ,
\;\;\;  r_{\infty}^3=r^3=36
\eeqa
\newpage

For this note first that by adjunction
\beqa
c(B_3)|_D=(1+D)\Big( 1+c_1(D)+c_2(D)\Big)|_D
\eeqa 
one has the relation
\beqa
\label{intersect} c_2(B_3)\cdot D= D^2 \cdot c_1(B_3)- D^3 +\int_D c_2(D) 
\eeqa
Further one has for the ${\bf P^1}$-fibration defined by $t$ that
(unspecified $c_i$ refer always to $B_2$)
\beqa
c_1(B_3)&=&2r+c_1+t\nonumber\\ c_2(B_3)&=&2c_1 r + c_1 t + c_2 
\eeqa
again from adjunction $c(B_3)=(1+c_1+c_2)(1+r)(1+r+t)$.
(\ref{intersect}) now gives $r^3=36$ as
\beqa
c_2(B_3)\cdot r&=&(-c_1t+c_2)r=-c_1t+3\nonumber\\
r^2 \cdot c_1(B_3)- r^3 +\int_r c_2(r) &=&-r(c_1t+t^2)+r^3+3
=-c_1t-36+r^3+3\nonumber
\eeqa

{\em The K\"ahler potential}

One can now compute the K\"ahler potential
for the K\"ahler form $J':=t_1L + t_2'r$. Alternatively and equivalently
one can consider $r_{\infty}$ instead of $r$
\beqa
J=t_1 L + t_2 r_{\infty}
\eeqa
One gets for the divisor volumes
$\tau_{r_{(\infty)}}=r_{(\infty)} J^2/2$ and for the total volume $V=J^3/6$
\beqa
\label{taus}
\tau_1:=\tau_r &=& \frac{1}{2}t_1^2 \;\;\;\; , \;\;\; \;\;\;
\tau_2:=\tau_{r_{\infty}} = \frac{1}{2}(t_1+6t_2)^2\nonumber\\
V&=&\frac{1}{6}(3t_1^2t_2+18t_1t_2^2+36t_2^3)
=\frac{1}{36}\Big( (t_1+6t_2)^3 - t_1^3 \Big)
=\frac{\sqrt{2}}{18}(\tau_2^{3/2}-\tau_1^{3/2})
\eeqa
implying that $\tau_2 > \tau_1$ in the K\"ahler cone.
With this one gets the K\"ahler potential
\beqa
\label{t_i potential} K(t_i)=-2\ln V
\eeqa

{\em Remark:} For direct comparison with the K\"ahler potential
(\ref{CY Kahler pot sigma L})
of the $N=2$ parent theory let us also give here the K\"ahler potential
w.r.t. $J'$ (i.e. $\sigma$ above becomes $r$ below)
\beqa
e^{-\frac{1}{2}K}=V=
\frac{1}{12}\Big(9(2t_2')^3-9(2t_2')^2 t_1+3(2t_2') t_1^2\Big)
\eeqa

\section{Integrating out $z_i$ and $\tau$ and working with $W_0$}

\resetcounter

The stabilisation of $\tau$ and the $z_i$ is achieved from
the conditions
\beqa
D_{i} W = \partial_{i} W + (\partial_{i}K)\, W  =  0 \;\;\; , \;\;\;
D_\tau W = \partial_{\tau} W + (\partial_{\tau}K)\, W  =  0\nonumber
\eeqa
In the given example one finds a stationary point 
for the values\footnote{with instanton corrections 
consistently ignored for $\Im z_1, \Im z_2 \geq 1$}
[\ref{DDF}]
$z_1=i= z_2\; , \tau = 3i$.
This vacuum of\footnote{With the first instanton correction 
one finds a near-by vacuum with $e^K|W|^2=1.379 \x 10^{-4}$.} 
$W=0$ is reached for the following flux combination
of $N_{flux}=352$ 
\beqa
\label{DDF fluxes}
(e_R^a \, | \, m_R^a)= (20, 0 , 0 \, | \, 0 , -69 , -28)
\;\;\; , \;\;\;
(e_{NS}^a \, | \, m_{NS}^a) = (0 , -4 , 0 \, | \, 49 , 18 , 6)\;\;\;\;\;
\\
W=69 z_1 + 28 z_2 + 20 F_0 
- \tau \Big( -49 -18 z_1 -6 z_2 -4 F_1 \Big)\;\;\;\;\;\;\;\;\;\;\;\;\;\;\;\;
\;\;\;\;\;\;\;\;\;\;\;\;\;\;\;\;\;\;\;\;\;\;\;\;\;\;\;\;\;\;\;\;\nonumber\\
= 30 z_1^3+30 z_1^2 z_2 + 10z_1z_2^2 + 154 z_1 + 58 z_2 + 40\xi 
+\tau\big( -2(3z_1+z_2)^2+ 27 z_1+9  z_2 +66 \big)\nonumber
\eeqa

\subsection{The K\"ahler stabilizing superpotential}

[\ref{DDF}] find no divisors leading to an 
instanton contribution in the superpotential following
the necessary condition $\chi(D)=1$ for the 
arithmetic genus [\ref{W}].\footnote{As actually $t=6l$ all $\chi$ values 
in (\ref{chi values}) are negative, 
so the flux influence can not change them to $1$.}
One computes $\chi(\pi^{-1}C)=-C^2 c_1(B_3)/2$ [\ref{G}].
Recall now that the base $B_3$ of $X_4$ was ${\bf P^1_z}$ fibered over 
$B_2={\bf P^2}$ and one had divisors $C=r$ and $r_{\infty}$ in $B_3$
as sections of this ${\bf P^1_z}$ fibration at $z=0$ and $z=\infty$, 
respectively.
For the projection $pr: B_3\ra B_2$ (with  $t=nl$, say) one finds 
for the $pr$-horizontal divisors (in $B_3$) 
$r$ and $r_{\infty}$ and for the $pr$-vertical divisor $pr^{-1}P$ 
(for $P=al$)
\beqa
\label{chi values}
\chi(\pi^{-1}r_{0/\infty})=\frac{1}{2}t(\pm c_1 - t)
=\frac{1}{2}n(\pm 3 - n)\;\;\; , \;\;\;
\chi(\pi^{-1}pr^{-1}P)=-P^2=-a^2
\eeqa

So one has to rely solely on gauge type divisors
(where contributions are induced from gaugino condensation;
cf. also [\ref{KV}]).
The concrete fourfold we work with following [\ref{DDF}]
has actually a $G_2$ (not $D_4$) singularity along $r$
and an $E_6$ or $E_8$ singularity along $r_{\infty}$.
This leads to values of $\chi(X_4)/24$ 
of $273, 129$, respectively (note however that $N_{flux}=352$).

With the volumes  $\tau_1:=\tau_r$ and $\tau_2:=\tau_{r_{\infty}}$ 
of the divisors $r$ and $r_{\infty}$ a superpotential 
\beqa
W=W_0 +C_1 e^{-2\pi a_1 \T_1}
+C_2 e^{-2\pi a_2 \T_2}
\eeqa
is generated where we introduced the complexified 4-cycle-volumes
$\T_j=\tau_j+i \th_j$ with $\th_i=-\int_{r_{(\infty)}}C_4$.
Assuming $W_0$ real (and $C_i=1$) and that the two 
exponential terms should equal approximately $W_0$
at the critical point, they should be also 
approximately equal there. In view of $\tau_1 < \tau_2$ this means
that $a_1 > a_2$. For some numerical solutions in the set-up
above, where $a_1=1/4$ and $a_2=1/12$ or $1/30$, with 
$W_0=10^{-5}$ or $10^{-30}$ cf.~[\ref{DDF}].

\newpage

\section{Integrating out the $z_i$ and working with $K_{eff}(\tau)$}

\resetcounter

The action for (gauge neutral) chiral mutiplets can be
expressed in terms of a single real function 
$G$. The scalar potential becomes 
(with $G_{i\bar\jmath}=\partial_i\partial_{\bar\jmath}G$ 
and $G_i=\partial_i G$, $G^i=G^{i\jb}G_\jb$) 
\beqa
\label{SUGRA V}
V=e^{G}\Big(G^{i\bar\jmath}G_i G_{\bar\jmath}-3\Big)
= e^K \Big(K^{i \overline{j}} D_i W \overline{D}_{\overline{j}} 
\overline{W} - 3W\overline{W}\Big)
\eeqa 
The index 
$i$ runs over all chiral superfields, here the moduli and the dilaton. 
$G$ can be, and usually is, (ambiguously) split as $G=K+\ln|W|^2$
into a real K\"ahler potential $K(z,\bar z)$ 
and a holomorphic superpotential $W(z)$. 
The supersymmetry conditions are $D_iW=W_i+K_i W=WG_i=0$. So, for $W\neq 0$,
the supersymmetry conditions are in terms of $G$ given by $G_i=0$.

{\em On the stability analysis in general}

\noindent
The stability analysis is based on the eigenvalues of the mass matrix. 
In the AdS case, one has to check whether the BF bound is satisfied. 
But in the case where one 
adds a term such that the vacuum energy becomes positive, one has to 
make sure that all eigenvalues of the mass matrix are actually positive. 
This must be done in a field basis where the kinetic terms are 
canonically normalized. However the eigenvalues in both bases are not the 
same but their signs are the same. 
If they are all positive in the original basis, they are also 
positive in the canonical basis 
(so a negative eigenvalue signals an instability). 

To parallelize the treatment of the stability criterion
of [\ref{Nilles}] let us recall also here the mass matrix, now in terms of $G$.
First, the stationarity condition $\p_k V=0$ becomes
\beqa
\p_k V=e^G\Big( (G^i G_i-3)G_k+(G^i\nabla_k G_i+G_k)\Big)
\eeqa
At the miminum one gets the $(kl)$-symmetric (as $G_{i\jb}$ is a 
K\"ahler metric) expression\footnote{For Minkowski vacua one is left with the 
first three terms (the last of these is missing in [\ref{WessB}]).}
\beqa\p_l\p_k V|_{dV=0}&=&
e^G\Big(
\nabla_l G_k+\nabla_k G_l+(G^i G_i-3)\nabla_l G_k+G^i\nabla_l\nabla_k G_i
+(G_l+G^i\nabla_l G_i)G_k\Big) \nonumber\\
&=&e^G\Big(\nabla_l G_k+\nabla_k G_l+G^i\nabla_l\nabla_k G_i
+(G^i G_i-3)(\nabla_l G_k-G_l G_k)\Big)
\eeqa

For the mixed derivatives one finds (using 
$\nabla_{\bar l}\nabla_k G_i=R_{\bar l k i}{}^j G_j$)\footnote{For Minkowski 
vacua one finds the first term and the last two terms, cf. [\ref{WessB}].}
\beqa
\p_{\bar l}\p_k V|_{dV=0}&=
e^G\Big( G_{k\bar l}+(G^i G_i-3)G_{k\bar l}
+G_k(G_{\bar l}+G^\jb\nabla_{\bar l}G_\jb)
+\nabla_k G_i\nabla_{\bar l} G^i+R_{\bar l k i}{}^j G_j G^i\Big)
\;\;\;
\eeqa

For a SUSY vacuum with $G_i=0$ one has the simple expressions
\beqa
\p_l\p_k V|_{\rm susy\,vacuum}&=&-e^G \p_k\p_l G\\
\p_{\bar l}\p_k V|_{\rm susy\,vacuum}&=&
e^G\Big( -2 G_{k\bar l}+G^{i\jb}\p_k G_i\p_{\bar l}G_\jb\Big)
\eeqa

\newpage

\subsection{Restricted preconsideration: the $(z_i,\tau)$-sector}

There are two equivalent
ways to handle the integrating-out procedure. Either one eliminates the 
$z_i$ in $G$ (cf. below) and then operates, after reinserting them, with 
a $G_{eff}$; alternatively one derives $W_{eff}^{non-holo}(\tau)$ 
from $D_i W_{flux}(z_i, \tau)=0$ first and takes then,
working in the $G$-formalism,  $W_{eff}^{hol}(\tau)=1$ 
and $K_{tot}(\tau)=K({\tau})+K(z_i(\tau))+\ln \, |W_{eff}^{non-holo}|^2$
and the corresponding $G_{eff}=K_{tot}$; for this 
cf.~[\ref{de Alwis}] and the example treated by us below.
Concretely from the expressions given earlier one gets then
the $z_i$-elimination conditions
\beqa 
\label{int out in z_i S sector}
D_{z_i} W=\del_{z_i}W + \del_{z_i}K \; W = 0 
\eeqa
These equations allow to eliminate the $z_i$
in $W_{flux}(z_i, \tau)$ and to get an effective superpotential
$W_{eff}^{non-holo}(\tau)$ (this replaces the earlier constant 
$W_0$ when also $\tau$ was integrated out)\footnote{the 
fluxes chosen earlier which lead to the concrete expressions 
for $W_{flux}$ gave a minimum at
$z_1=z_2=i, \tau=3i$ in the procedure where $\tau$ is also
integrated out; we can, of course, choose other fluxes.}.
One gets then, in this seond way, 
the complete K\"ahler potential in $\tau$ as follows
\beqa
K(\tau)&=&-\ln(\tau - \bar{\tau})\\
K_{eff}(\tau)&=&K(z_i(\tau))\\
K_{tot}(\tau)&=&K(\tau)+K_{eff}(\tau)+\ln|W_{eff}^{non-holo}(\tau)|^2
\eeqa

Concerning the first procedure, 
as explained in the introduction, our strategy will be to 
solve the supersymmetry conditions for all the fields which appear in 
$G$. We first eliminate the $z_i$ by solving, 
for a suitable choice of fluxes, 
$\partial_{z_i}G=0$ where 
\beqa
G=K(\tau)+K(z_i)+\ln |W_{flux}(z_i, \tau)|^2
\eeqa
The equations can be solved for the $z_i$ and the solutions, which depend 
on $\tau$, produce $G_{eff}(\tau)$ when reinserted into $G$. 
Two points are worth mentioning here. First, since we are working with 
the real function $G$, questions of holomorphicity of the superpotential
do not arise. This is an issue, since the SUSY conditions $D_{z_i}W=0$ are 
not holomorphic and would lead to a non-holomorphic $W_{eff}$. 
Secondly, one also has to insert the solutions $z_i(\tau)$ 
into $K(z_i)$ and not merely into $W_{flux}$.

Note that, for $W\neq 0$, both procedures are equivalent as the respective
elimination conditions are $G_i=0$ and $D_iW=0$.

Before we come to the main issue of this note, the consideration of the 
case with two K\"ahler parameters, we give first the procedure in the
generic one parameter case, following [\ref{Nilles}] 
but including $K(z_i(\tau))$ and $\ln|W_{eff}^{non-holo}(\tau)|^2$.
We will see that including the $\T$-sector leads to a slight change
in the procedure.

\subsection{\label{full elim subsect}
Full $(z_i, \tau, \T)$-sector: the generic one parameter case}

In the generic set-up with one K\"ahler parameter one would start
in analogy to [\ref{Nilles}] with
\beqa
W=W_{flux}(z_i, \tau)+W^{(\T)}=A(z_i)+\tau B(z_i)+Ce^{-a\T}
\eeqa
with the K\"ahler potential (where $K(\T)=-2\ln (\T +\bar{\T})^{3/2}$)
\beqa
K=-\ln (\tau - \bar{\tau})+K(z_i)+K(\T)
\eeqa
Integrating out the $z_i$ by $G_i = 0$ resp. 
$G_i \cdot W = D_i W = D_i (W_{flux}+Ce^{-a\T}) = 0$ 
gives
\beqa
D_i W = A_i+\tau B_i+K_i(W_{flux}+Ce^{-a\T}) = 0
\eeqa
where compared with the previous (\ref{int out in z_i S sector}) 
the exponentially suppressed perturbation term
$K_iW^{(\T)}$ has arisen. Therefore now the previous $z_i=z_i(\tau)$ 
becomes here 
$z_i=z_i(\tau,\T)
=z_i(\tau)+\delta W^{(\T)}+\epsilon \bar{W}^{(\bar{\T})} +\dots $, 
cf. [\ref{de Alwis}].
So one is led to use a $W_{eff}^{holo}(\tau, \T)\equiv 1$ 
and the K\"ahler potential
\beqa
\label{G true}
G_{eff}=K_{eff}
&=&-\ln (\tau-\bar{\tau})+K\big(z_i(\tau, \T)\big)
-3\ln (\T + \bar{\T})\nonumber\\
&&+\ln|A\big(z_i(\tau, \T)\big)+\tau B\big(z_i(\tau, \T)\big)+Ce^{-a\T}|^2
\eeqa
Note that to get $G_{eff}$ in (\ref{G true}), it would
have been again equivalent (for $W\neq 0$) to start with the full 
$G^{full}=-\ln (\tau - \bar{\tau})+K(z_i)+K(\T)
+\ln|W_{flux}(z_i, \tau)+W^{(\T)}|^2$, 
solve there $G^{full}_i=0$ for the $z_i(\tau,\T)$ 
and reinserting into $G^{full}$, for one has 
$D_i W_{flux}=WG^{full}_i$.

Note in passing that the procedure described 
is not the same as doing the previous 
integrating-out procedure just in the $(z_i, \tau)$-sector with
$W_{eff}^{holo}(\tau)\equiv 1$ and 
just adding the $\T$-sector.
This leads to the following point: in the two-step analysis of KKLT
all moduli of $W_{flux}(z_i, \tau)$ were integrated out; this
left behind just a $W_0$ when afterwards the further $W^{(\T)}$ was included. 
However, as [\ref{Nilles}] pointed out, this procedure may not be justified 
as the masses of $\tau$ and $\T$ can be comparable; so in [\ref{Nilles}] 
just the $z_i$ were integrated out, leaving a $W_{eff}(\tau)$ behind instead 
of the constant $W_0$. This procedure, as outlined in [\ref{de Alwis}] and 
described also below, has to be supplemented, at least, with the inclusion of 
$K(z_i(\tau))$ and a proper treatment of $W_{eff}^{non-holo}(\tau)$. 
As we described above, even including these points, there would be still a
remnant of the linkage of $\tau$ with the $z_i$, if $z_i(\tau)$ 
would be assumed to be just a function of $\tau$ (what would arise from
integrating it out while considering just $W_{flux}(z_i,\tau)$, and including
$W^{(\T)}(\T)$ only afterwards) instead of determining a $z_i(\tau, \T)$ from
the full $W=W_{flux}(z_i,\tau)+W^{(\T)}(\T)$.

\newpage

\section{The two-parameter example}

\resetcounter

Take the flux components with
the general full magnetic sector
\beqa
(e_R^a \, | \, m_R^a)= (0, g , h \, | \, -e , -a , -b)
\;\;\; , \;\;\;
(e_{NS}^a \, | \, m_{NS}^a) = (0 , 0 , 0 \, | \, e' , c , d)
\eeqa
Concerning the two special magnetic fluxes $e$ and $e'$ which contribute
in $W_{flux}$ only by adding a constant $E=e+e'\tau$, which is independent of
the complex structure moduli, we will set at first $e=e'=0$.
We will employ the following notation (and $z_k=x_k+iy_k$)
\beqa
A:=a+\tau c=A_1+iA_2\;\;  & , & \;\;  B:=b+\tau d=B_1+iB_2 \\
Z:=3z_1+z_2=X+iY\;\; & , & \;\; y:= y_2 \\
p:=\frac{a-3b}{3} \;\; &  , & \;\; q:=\frac{c-3d}{3}\\
W^{(\T)}= & \sum_{k=1}^2 W^k & = \sum_{k=1}^2 C_k e^{-a_k\T_k}
\eeqa
We will always stick to the case $q=0$, i.e. $c= 3d$
(note that this was also satisfied 
in the flux-choice (\ref{DDF fluxes})); we exclude the case $c=0=d$
where things collapse trivially.

We will then write the two SUSY conditions $D_{z_i}W=0$ which allow us
to eliminate the $z_i$ and write $W_{eff}(\tau)$. The way we proceed to do this
in practice amounts to writing $W_{flux}$ 
(with the help of the SUSY conditions) purely as function of $\tau$ 
and only one of the four real variables $X,Y,x_2,y$, concretely either $Y$ 
or $y$, where the latter furthermore depends 
in a simple polynomial way on $\tau$.

We then proceed and
include the $\T_k$-sector. For that purpose we generalize the above set-up
first to include fluxes $m_R^0=-e, m_{NS}^0=e'$ and the corresponding
contribution $E=e+e'\tau=E_1+iE_2$ in $W_{flux}$. 
The latter will be used then to mimic the $W^{(\T)}$.
This will be described more fully below in our main example, the case II.
In total the following cases will be described
\beqa
\un{{\bf CASE \;\; I}}\;\;\;\;\;\;\;\;\;\;\;\;\;\;\;\;\;\;\;\;\;\;\;\;
\;\;\;\;\;\;\;\;\;\;\;\;\;\; g=0 \; , \;\;\;\;\; h=0
\nonumber\\
\un{{\bf CASE \;\; II}}\;\;\;\;\;\;\;\;\;\;\;\;\;\;\;\;\;\;\;\;\;\;\;\;
\;\;\;\;\;\;\;\;\;\;\;\;\;\; g\neq 0 \; , \;\;\;\;\; h=0
\nonumber\\
\;\;\;\;\;\;\;\;\;\;\;\;\;\;\;\;\;\;\un{{\bf CASE \;\; III}}
\;\;\;\;\;\;\;\;\;\;\;\;\;\;\;\;\;\;\;
\;\;\;\;\;\;\;\;\;\;\;\;\;\;\;\;\;\; g\neq 0 \; , \; h=-3g\nonumber
\eeqa
We will get an explicit expression for $W_{eff}$ in all 3 cases. 
In the case II, which constitutes our main example, we will proceed
and analyse the SUSY conditions completely in closed analytical form
and show that no SUSY solution exists. As the main issue here is 
the explicit derivation of $W_{eff}$ and the concrete analytical
treatment of the ensuing SUSY conditions we give this in some detail so that 
one can follow the line of derivation.

\newpage
\subsection{The case I: $g=h=0$}

We will at first assume that $g=h=0$. 
Although this leads to the somewhat degenerate case of $L=0$
we mention this case first as some equations appear here first 
in a somewhat simpler form than in our main example case II.
This leads to 
\beqa
W_{flux}=az_1+bz_2+\tau(cz_1+dz_2)=Az_1+Bz_2=\frac{A}{3} Z-pz_2
\eeqa
which gives as elimination equations (where $z_k=x_k+iy_k$)
\beqa
\label{first}
D_{z_1}W=0\Longrightarrow 
2i\frac{A}{3} =\frac{\frac{1}{3}(3y_1+y_2)^2}{y_1(3y_1^2+3y_1y_2+y_2^2)+i\xi}
\Big( Az_1+Bz_2\Big)\\
\label{second}
D_{z_2}W=0\Longrightarrow 
2iB =\frac{y_1(3y_1+2y_2)}{y_1(3y_1^2+3y_1y_2+y_2^2)+i\xi}
\Big( Az_1+Bz_2\Big)
\eeqa
This yields (for $Y\neq 0$, as is needed to keep things nontrivial) 
the compatibility equation
\beqa
\label{the one complex equation}
\frac{A}{3}\, (Y^2-y^2)=B Y^2
\eeqa
As the $y_k$ are real and $A_2/3=B_2$
the imaginary part here gives $y=0$, 
with $pY^2=0$ as remaining relation; one gets therefore also
$p=0$ and the expressions $W_{flux}=\frac{A}{3} Z$ and
\beqa
\label{expression for Wflux}
W_{flux}=2i\frac{A}{3}\; \frac{3y_1^3+i\xi}{3y_1^2}
\eeqa
The resulting relation $Z=2i\frac{3y_1^3+i\xi}{3y_1^2}$ gives $X=0$ 
and $Y=2\frac{Y^3+9i\xi}{3Y^2}$ or
\beqa
Y&=&\sqrt[3]{2\cdot 9 i\xi}\;\;\;\;\; \;\;\;
\big( \Longleftarrow \;\; Y^3+3CY^2 -2\cdot 9i\xi=0
\;\;\; \mbox{with} \;\; C:= 0\big)\\
W_{eff}(\tau)&=&\frac{A}{3} Z = iAy_1 = i(a+c\tau)\sqrt[3]{\frac{2}{3}i\xi}
\eeqa

(Note that, if one would force now also the third condition 
(usual KKLT procedure)
\beqa
\label{KKLT procedure integrating out}
D_{\tau}W_{flux}=0\Longrightarrow 2idZ=\frac{1}{\tau_2}\Big( Az_1+Bz_2\Big)
\eeqa
one would get $2id\tau_2=\frac{a+c\tau}{3}$, i.e. $\tau= -\frac{a}{c}$
with $\tau_2=0$, i.e. no (physical) solution.)

Although this example is somewhat degenerate ($L=0$) let us proceed and
include the $\T_k$-sector. So we generalize the above 
to include fluxes $m_R^0=-e, m_{NS}^0=e'$ and the corresponding
contribution $E=e+e'\tau=E_1+iE_2$ in $W_{flux}$. Then the 
trivial $C$ above changes to an $E$-dependent $\C$ (we use now the 
abbreviated notation $\cA:=A/3$).

\newpage
\noindent
One gets $W_{flux}=\cA Z+E$ and equating with 
(\ref{expression for Wflux}) gives for the real and imaginary part
\beqa
X&=&-\frac{\cA_1E_1+\cA_2E_2}{\cA_1^2+\cA_2^2}\\
\label{cubic equation case I}
Y^3+3\C Y^2-2\cdot 9i\xi&=&0\;\;\;\;\;\;\;\;
(\C:=\frac{\cA_1E_2-\cA_2E_1}{\cA_1^2+\cA_2^2})
\eeqa
With this cubic determination equation one gets for the
effective K\"ahlerpotential (up to an additive constant) 
and the effective superpotential (with $E:= W^{(\T)}$) and $G$-function
\beqa
K_{eff}(\tau, \T_k)&=&-\ln(Y^3-y^3+9i\xi)=-2\ln Y -\ln \cB\;\;\;\\
W_{eff}(\tau, \T_k)&=&\cA(X+iY)+E
=\frac{2i}{3}\, \cA\, \frac{Y^3+9i\xi}{Y^2}=i\cA\cB\\
G_{eff}(\tau, \T_k)&=&K_{eff}(\tau, \T_k)+\ln |W_{eff}(\tau, \T_k)+W^{(\T)}|^2
-\ln(\tau-\bar{\tau})+K(\T_k)
\eeqa
Here we use the following notation
\beqa
\cB = Y + \C\;\;\; ,  \;\; s_k=(-1)^k\frac{3}{2}
\frac{t_k^{1/2}}{t_2^{3/2}-t_1^{3/2}}
\;\;\;\;\;\;\;\;\;\;\;\;\;\;\;\;\;\;\;\;\;\;\;\;\;\;\;\;\;\;\;\;\;\;\;\\
\Gamma = \cB\frac{\cA \bar{\cA}\cB+\big(\cA_1\Im\, W - \cA_2\Re\, W\big)}
{\big(\cA_1\cB+\Im\, W\big)^2+\big(-\cA_2\cB+\Re\, W\big)^2}\; , \;\;
\Gamma'=\frac{-i\bar{\cA}\cB+\ov{W^{(\T)}}}
{\big(\cA_1\cB+\Im\, W\big)^2+\big(-\cA_2\cB+\Re\, W\big)^2}\;\;\;
\eeqa
Note that one still has $y=0$ as before. Furthermore $Y\neq 0 $ and
the effective K\"ahler potential gives from $Y^3-y^3+9i\xi=Y^3+9i\xi=
\frac{3}{2}Y^2(Y+\C)$ that also $\cB\neq 0$.

Furthermore one has the following relations ($l$ stands for $k$ or $\tau$)
\beqa
Y_l(Y+2\C)&=&-\C_lY\\
\C_k=3i\frac{a_kW^k}{2A}\; \;\; & , & \;\;\;
\C_{\tau} = 3i\frac{cW^{(\T)}}{2A^2}\\
Y_l = \C_l \; \frac{-Y}{Y+2\C}\;\; & , & \;\;
 \, \cB_l =  \C_l \;\frac{2\C}{Y+2\C}
\eeqa

From $G_{eff}$ (in the following denoted by $G$)
one gets as SUSY conditions
\beqa
G_k&=&-2\frac{Y_k}{Y}+(2\G-1)\frac{\cB_k}{\cB}-a_kW^k\, \G'+s_k=0\\
G_{\tau}&=&-2\frac{Y_{\tau}}{Y}
+(2\G-1)\frac{\cB_{\tau}}{\cB}+i\frac{c}{3}\cB\, \Gamma'+\frac{i}{2\tau_2}=0
\eeqa
For a full treatment of such 
a set of equations we refer to the case II below. As the case I 
is somewhat degenerate ($L=0$) we are content here to derive $W_{eff}$
(for more details on the eliminations following from the detailed treatment
of the SUSY equations, along the lines of the complete procedure
as in case II, cf. the appendix).

\newpage

\subsection{The case II: $g\neq 0 , h=0$}

\noindent
For an example where $L=cg/2\neq 0$ take
(again in the case $A_2=3B_2$, i.e.\footnote{We assume $c\neq 0 \neq d$
to retain a proper $\tau$-dependence in $W_{flux}$ after integrating out 
the $z_i$.} 
$c=3d$)
\beqa
(e_R^a \, | \, m_R^a)= (0, g , 0 \, | \, 0 , -a , -b)
\;\;\; , \;\;\;
(e_{NS}^a \, | \, m_{NS}^a) = (0 , 0 , 0 \, | \, 0 , c , d)
\eeqa
which gives (with $A=A_1+iA_2:=a+c\tau , B=B_1+iB_2:=b+d\tau $)
\beqa
W_{flux}=gF_1+ az_1+bz_2+\tau(cz_1+dz_2)
=g \Big(-\frac{1}{2}Z^2 + \frac{3}{2}Z + \frac{17}{4} \Big)+Az_1+Bz_2
\eeqa
We use the following abbreviations (for $g\neq 0$)
\beqa
\label{definitions}
C:=\frac{A_1 + \frac{9}{2}g}{3p}\;\; , \;\;\;\;\;\;\;
\beta=\frac{B_2}{g}
\eeqa
Useful relations are
\beqa
Az_1+Bz_2&=&\frac{A}{3}Z-pz_2\\
9y_1(3y_1^2+3y_1y_2+y_2^2)&=&Y^3-y_2^3
\eeqa
One gets as elimination equations $D_{z_1}W=0$ and $D_{z_2}W=0$ 
\beqa
\label{first 2}
\Big[ \frac{A}{3} - g \big( Z-\frac{3}{2} \big) \Big]
&=&\frac{3}{2i}\frac{Y^2}{Y^3-y^3+9i\xi}
\Big( gF_1+Az_1+Bz_2\Big) \\
\label{second 2}
\Big[B - g \big( Z-\frac{3}{2} \big) \Big]
&=&\frac{3}{2i}\frac{Y^2-y^2}{Y^3-y^3+9i\xi}
\Big( gF_1+Az_1+Bz_2\Big) 
\eeqa
(recall we denote $y_2$ just by $y$).
Now the imaginary and real part of the compatibility equation
\beqa
\label{compatibility equation}
\Big[ \frac{A}{3} - g \big( Z-\frac{3}{2} \big) \Big]
(Y^2-y^2)=
\Big[B - g \big( Z-\frac{3}{2} \big) \Big]
Y^2
\eeqa
yield for $y\neq 0$ and $g\neq 0$ the eliminations
\beqa
\label{X Y eliminations}
Y=\beta\;\;\;\; , \;\;\;\;
X=\frac{p}{g}\Big( C -\frac{\beta^2}{y^2}\Big)
\eeqa
Finally we have 
to solve the imaginary and real part of (\ref{first 2}), say,
which gives, noting that the big fraction on the right hand side is real 
and that (by explicit evaluation)
\beqa
\Im \; W_{flux} = p\big(  C\beta-y\big)
\eeqa
finally
%\footnote{$\Re \; W_G=0$ from (\ref{first 2}) gives
%$x_1=\frac{g}{A_1-3B_1}\Big( \frac{1}{2}X^2
%-\big( \frac{3}{4}+\frac{B_1}{g}\big)X
%+\big( \frac{1}{2}\beta^2-\frac{17}{4}\big)\Big)$}
the cubic equation
\beqa
\label{cubic y_2}
y^3-3C\beta y^2
+2\big(\beta^3+9i\xi\big)=0
\eeqa

Let us remark on the cubic equation occuring here.\\
1) (\ref{cubic y_2}) is of course easily solved in general. If 
one would be interested only in purely imaginary $\tau$
(note that  $\tau_2=s=Re\, S$ under $\tau =: i S$, 
where $\tau=\tau_1+i\tau_2$), 
i.e. $A_1=a, B_1=b$,
and furthermore in the special case 
$C=0$, i.e. $a=-\frac{9}{2}g$, then
\beqa
W_{eff}(\tau_2)=-ipy=i(b+\frac{3}{2}g)
\Big( -2\big(\frac{d^3}{g^3}\tau_2^3+9i\xi\big)\Big)^{1/3}
\eeqa
2) In the general case one gets for 
$r:=y-C\beta$ with $W_{eff}=-ipr$ the equation 
\beqa
\label{Cardano}
r^3+Pr+Q=r^3-3C^2\beta^2 r + 2\Big( (1-C^3)\beta^3+9i\xi\Big)=0
\eeqa
(note $\beta, C, i\xi, P,Q \in {\bf R}$)
with the three solutions given by the Cardano formula
%\beqa
%\label{Cardano solutions}
%r_i=\om^i \sqrt[3]{-\frac{Q}{2}+\sqrt{\big(\frac{Q}{2}\big)^2
%+\big(\frac{P}{3}\big)^3}}
%+ \om^{2i}\sqrt[3]{-\frac{Q}{2}-\sqrt{\big(\frac{Q}{2}\big)^2
%+\big(\frac{P}{3}\big)^3}}
%\eeqa
where one solution is always real and all three are real for
$0\geq D:=(Q/2)^2+(P/3)^3$. 

Now let us come back to the use of this information for our main goal,
whic is to derive the supersymmetry conditions from the effective $G$-function.
One now has a $K_{eff}(\tau)=K(z_i(\tau))$ which can be written 
in various equivalent ways
\beqa
K_{eff}(\tau)&=&
-\ln \Big( \beta^3+9i\xi-y^3
\Big) -\ln \frac{4i}{9}\\
&=&-\ln \Big( \beta^3+9i\xi-C\beta y^2
\Big) -\ln \frac{4i}{3}\\
\label{K_eff}
&=&-\ln y^2 -\ln (y-C\beta)-\ln \frac{-2i}{3}
\eeqa
giving in total (up to the constant $-\ln \frac{-2i}{3}$)
\beqa
K_{tot}(\tau)=-\ln (\tau-\bar{\tau})-2\ln y -\ln (y-C\beta)
\eeqa
Note that, in contrast to the original piece $-\ln (\tau-\bar{\tau})$,
the new, full $K_{tot}(\tau)$ depends not just on $\tau_2$ as the coefficient
$C$ (which determines together with $\beta$ the $y$) depends on $\tau_1$. 

Because of its non-holomorphicity one gets
(for $r\in {\bf R}$) from $W_{eff}$ where
\beqa
W_{eff}(\tau)=ip(\beta C - y)
\eeqa
the further contribution (up to the constant $2\ln p$)
\beqa
\ln |W_{eff}^{non-holo}(\tau)|^2 = 2\ln (y-C\beta) 
\eeqa

Up to now we have followed the treatment just in the $(z_i, \tau)$-sector
(cf. our discussion in the subsection on the corresponding
restricted preconsideration). Now we have to properly enhance these
results by going to the consideration of the full $(z_i, \tau, \T_k)$-sector.

\newpage

{\em Including the $\T_i$-sector}

Up to now we have expressed the $z_i$ as functions of $\tau$ 
from $D_i W_{flux}=0$. According to the discussion in sect. 
\ref{full elim subsect} we want actually to obtain $z_i(\tau, \T_i)$
from $D_i (W_{flux}+W^{(\T)})=0$. For this we will 
1) give the explicit expressions $z_i(\tau)$,
2) generalize our flux-example to include a constant flux and give there the
$z_i(\tau)$,
3) interpret {\em formally} the additional term $W^{(\T)}$, 
which is constant in $\tau$, as such an additional flux term  and
4) give thereby the full $z_i(\tau, \T_i)$.

1) Note first that $\mbox{Re} W_{flux}=0$ from (\ref{first 2}) gives
\beqa
x_1=\frac{g}{A_1-3B_1}\Big( \frac{1}{2}X^2+\frac{1}{2}\beta^2
-(\frac{3}{2}+\frac{B_1}{g})X-\frac{17}{4}\Big)
\eeqa
Therefore the two variables $Z=3z_1+z_2=X+iY$ and $z_1=x_1+iy_1$ 
are completely determined as functions of $\tau$ and $y=y_2$
(and therefore as functions of $\tau$ alone): for $X$ and $Y$
this was given above, and for $z_1$ it is, together with $y_1=(Y-y_2)/3$,
given here.

2) Now let us generalize this set-up by including an additional 
constant (in the $z_i$)
contribution in $W_{flux}$. For this we start with the
enhanced flux vector 
\beqa
(e_R^a \, | \, m_R^a)= (0, g , 0 \, | \, -e , -a , -b)
\;\;\; , \;\;\;
(e_{NS}^a \, | \, m_{NS}^a) = (0 , 0 , 0 \, | \, e^{'} , c , d)
\eeqa
which gives (with $E=e+e^{'}\tau=E_1+iE_2$) the flux-superpotential 
$W_{flux}^{'}=W_{flux}+E$ 
\beqa
W_{flux}^{'}=gF_1+Az_1+Bz_2+E
\eeqa
In this generalized situation (with $p\tilde{E}=E$)
the elimination of $y$ reads now
(the expressions (\ref{X Y eliminations}) for $X$ and $Y$, coming 
from (\ref{compatibility equation}), have still just $g$ and $C$ and
remain the same)
\beqa
\label{tilde cubic}
y^3-3(\beta C + \tilde{E}_2)y^2+2(\beta^3+9i\xi)=0
\eeqa
Formally one could even absorb the $\tilde{E}_2=:\tilde{e}\tau_2=pe^{'}\tau_2$ 
in a $\tilde{C}$ built with an $\tilde{g}:=g(1+\frac{2e^{'}}{3d})$, 
i.e., $\beta C + \tilde{E}_2=\beta \tilde{C}$
(not to be used in (\ref{X Y eliminations})).
Finally one gets the shift $x_1^{'}=x_1-\frac{E_1}{a-3b}$.

3) Now the elimination condition for $z_i=z_i(\tau, \T_i)$
\beqa
\del_i W_{flux}+K_i\big( W_{flux}+W^{(\T)}\big) =0
\eeqa
can formally be interpreted as the elimination condition
$\del_i W_{flux}^{'} + K_i W_{flux}^{'} =0$ for $z_i=z_i(\tau)$
with respect to an enhanced flux-superpotential 
$W_{flux}^{'}=W_{flux}+E$ 
where the latter will be interpreted formally as $W^{(\T)}$.

4) Alltogether we get thereby the searched for expressions $z_i(\tau, \T_i)$
as the $z_i^{'}(\tau)$ in the previous set-up, generalized by 
putting formally $E:=W^{(\T)}$.

\subsubsection{The SUSY conditions for the K\"ahler moduli $\T_i$ and 
the dilaton $\tau$}

Let us now apply the previous considerations 
concerning $G_{eff}$ in (\ref{G true})
to our two-parameter model. In the end we will discuss stable minima
for $V$ in (\ref{SUGRA V}). One finds, for\footnote{i.e., the $y$
in (\ref{tilde cubic}) can be understood as the ordinary $y$ in
(\ref{cubic y_2}) with $C\ra \tilde{C}$}
$y$ in (\ref{tilde cubic}),
with the help of (\ref{K_eff}) that (up to an additive constant)
\beqa
K_{eff}(\tau, \T_i)=
K\big(z_i(\tau, \T_i)\big)=K\big(z_i^{'}(\tau)\big)
=-\ln y^2 - \ln (y-\beta \tilde{C})
\eeqa
Similarly one has with $E=W^{(\T)}$, $E=p\tilde{E}$ 
and $\beta C + \tilde{E}_2=\beta \tilde{C}$
where $\tilde{C}$ is a function not only of $\tau_1$, 
as in (\ref{definitions}), but also of the  $\T_i$
(with $W^{(\T)}(\T_i)=\sum C_k e^{-a_k \T_k}$)
\beqa
\label{tilde C}
\tilde{C}=\frac{a+c\tau_1 + \frac{9}{2}g
+\frac{3g}{d\tau_2}\mbox{Im}W^{(\T)}(\T_i)}{a-3b}
\eeqa
(for simplicity we will assume $C_i\in {\bf R}$)
the following expression
\beqa
W_{eff}(\tau, \T_i)=
W_{flux}\big(z_i(\tau, \T_i), \tau\big)&=&W_{flux}\big(z_i^{'}(\tau)\big)
=i\Big[ p(\beta C - y)+\mbox{Im}W^{(\T)}\Big]\nonumber\\
&=&i p\big(\beta \tilde{C} - y\big)
\eeqa
Therefore we find alltogether for the $G_{eff}$ in (\ref{G true})
(up to an additive constant)
\beqa
G_{eff}&=&-\ln(\tau - \bar{\tau})+K_{eff}\big(\tau, \T_i\big)
+\ln\Big|W_{eff}\big(\tau, \T_i \big)+W^{(\T)}(\T_i)\Big|^2
+K(\T_i)\\
&=&-\ln(\tau - \bar{\tau})-2\ln y - \ln (y-\beta \tilde{C})
+\ln\Big|ip(\beta \tilde{C} - y)
+\sum_{i=1}^2 C_i e^{-a_i\T_i}\Big|^2\nonumber\\
&&-2\ln\Big( (\T_2+\Tb_2)^{3/2}-(\T_1+\Tb_1)^{3/2}\Big)
\eeqa
from which one gets the three $SUSY$ conditions $G_i=0$.
When searching for solutions one has to satisfy also $L=N_{flux}=cg/2
\leq e(X_4)/24=273$ and $129$ for $a_2=1/12$ and $1/30$, respectively.
(When using $W^{(\T)}$ numerically 
one has to make the shift $a_i\ra 2\pi a_i$.)

Using $\beta = \frac{d}{g}\tau_2,$ and
$\beta \C=\beta C + \frac{1}{p}\mbox{Im}W^{(T)}$
one has 
\beqa
\beta_{\tau}=\frac{d}{2ig}  \;\;\; & , & \;\;\;
(\beta\C)_{\tau}=\beta_{\tau}C+\beta \frac{c/2}{a-3b}
%=\frac{d}{2ig}\frac{a+c\tau +\frac{9}{4}\tilde{g}}{a-3b}
\\
\beta_{\T_i}=0
\;\;\; & , & \;\;\;
(\beta\C)_{\T_k}=\frac{i}{2p}\, a_k\, C_ke^{-a_k \T_k}
\eeqa
From (\ref{tilde cubic}) one finds ($l$ is $\tau$ or $\T_i$)
\beqa
\label{cubic diff}
y_{l}(y-2\beta\C)
=(\beta\C)_{l}y-2\frac{\beta_{l}\beta^2}{y}
\eeqa

{\em Evaluating the SUSY conditions}

For evaluation of the SUSY conditions we use the following abbreviations 
\beqa
\cB=\beta \tilde{C}-y\;\;\; , \;\;\; 
s_k = - (-1)^k \, \frac{3}{2} \, \frac{t_k^{1/2}}{t_2^{3/2}-t_1^{3/2}}
\;\;\; , \;\;\;
W^{T}=\sum_{k=1}^2 W^k=\sum_{k=1}^2 C_k e^{-a_k(t_k+i\th_k)}\\
\G = p \cB \; \frac{p\cB + Im \, W^{T}}{(p\cB + Im \, W^{T})^2+(Re\, W^{T})^2}
\;\;\; , \;\;\;
\G' = \frac{-i(p\cB + Im \, W^{T})+Re\, W^{T}}
{(p\cB + Im \, W^{T})^2+(Re\, W^{T})^2}
\eeqa
($\cB, s_k, \G$ are real quantities). 
One gets for the SUSY conditions
\beqa
G_k&=&-2\frac{y_k}{y}+(2\G-1)\frac{\cB_k}{\cB}-a_kW^k\G'+s_k=0\\
G_{\tau}&=&-2\frac{y_{\tau}}{y}
+(2\G-1)\frac{\cB_{\tau}}{\cB}+\frac{i}{2\tau_2}=0
\eeqa
Note that $y\neq 0$ or else one would have 
$\beta^3+9i\xi=0$ by the cubic equation
and $K(z_i)=\ln\big( Y^3-y^3+9i\xi \big)
=\ln\big(\beta^3 -y^3+9i\xi\big)$ would degenerate. Note furthermore
that $y-2\beta \C\neq 0$ or else (\ref{cubic diff}) would give 
$(\beta\frac{c/2}{a-3b}+\frac{d}{2ig}C)y^2=2\frac{d}{2ig}\beta^2$, i.e.
$c=0=d$ from the real part, violating our assumption. Applying
(\ref{cubic diff}) to $l=k$ likewise gives the immediate contradiction $C_k=0$.

For the following it is useful to collect some expressions
(subscript $k$ indicates $\del_{\T_k}$)
\beqa
(\beta \C)_k&=&\frac{i}{2p}\; a_kW^k\\
\label{k derivatives real}
y_k = (\beta \C)_k \; \frac{y}{y-2\beta \C} \;\; & , & \;\;\;
\cB_k = - \, (\beta \C)_k \; \frac{2\beta \C}{y-2\beta \C}
\eeqa
As the splitting in real and complex parts will be crucial below
note that here the fractions are real so that the real and imaginary parts
of $y_k$ and $\cB_k$ go with the corresponding real and imaginary parts 
of $(\beta \C)_k $ (times the fraction).

Similarly one gets in the $\tau$-sector
\beqa
(\beta \C)_{\tau}&=& \beta \; \frac{c/2}{3p} - i\,  \frac{d}{2g}\; C\\
Re\, y_{\tau} = Re \, (\beta \C)_{\tau} \; 
\frac{y}{y-2\beta \C} \;\; & , & \;\;\;
Re \, \cB_{\tau} = -Re \, (\beta \C)_{\tau} \; \frac{2\beta \C}{y-2\beta \C}\\
Im\, y_{\tau} = -\frac{d}{2g}\; \frac{C y^2-2\beta^2}{y(y-2\beta \C)} 
\;\; & , & \;\;\;
Im \, \cB_{\tau} = -\frac{d}{2g}\; \frac{2\beta^2-2\beta\C C y}{y(y-2\beta \C)}
\eeqa

\newpage

Now let us consider the real and imaginary parts of $G_k=0$ and $G_{\tau}=0$
\beqa
\label{real G_k}
-2\frac{Re\, y_k}{y}+(2\G-1)\frac{Re\, \cB_k}{\cB}
-a_k\frac{Re\, \big[ W^k (-ip\cB +\bar{W})\big]}{|ip\cB+W|^2}&=&-s_k\\
\label{imag G_k}
-2\frac{Im\, y_k}{y}+(2\G-1)\frac{Im\, \cB_k}{\cB}
-a_k\frac{Im\, \big[ W^k (-ip\cB +\bar{W})\big]}{|ip\cB+W|^2}
&=&0\\
\label{real G_tau}
-2\frac{Re \, y_{\tau}}{y}+(2\G-1)\frac{Re\, \cB_{\tau}}{\cB}&=&0\\
\label{imag G_tau}
-2\frac{Im \, y_{\tau}}{y}+(2\G-1)\frac{Im\, \cB_{\tau}}{\cB}
&=&-\frac{1}{2\tau_2}
\eeqa
(here $W:=W^{T}$). Let us draw some consequences 
(we had $C_i\in {\bf R}$ for simplicity)
\beqa
\label{basic relation}
2\cB&=&-2(2\G -1)\, \beta \C \;\;\;\; \;\;\;\;\;\;\;\;\;\;\;\;\;
\;\;\;\;\;\;\;\;\;\;\;\;\;\;\;\;\;\;\;\; \;\;\;\;\;\;\;\;
(\Longrightarrow y=2\G \beta \C)\\
\label{cons 1}
Re\, W^k&=&0 \;\;\;\; \Longrightarrow \;\; \th_k=\pm \frac{\pi}{a_k}
\;\;\; \mbox{and} \;\;
Re\, W=0 \;\;\;\; \;\;\;\;\;\;\;\;
(\Longrightarrow\; \Gamma=\frac{p\cB}{p\cB + Im \, W})\\
\label{cons 2}
\frac{s_k}{a_k}&=&\frac{Im\, W^k }{p\cB + Im \, W}
=\frac{\pm C_k e^{-a_k t_k} }{p\cB + \sum \pm C_j e^{-a_j t_j} }
\;\;\;\;\;\;\;\;\; (\Longrightarrow\; 
\sum \frac{s_k}{a_k}=\frac{Im\, W }{p\cB + Im \, W})\;\;\;\;\;\;\;\;\\
\label{cons 4}
\C y^2&=&-2\beta^2\;\;\;\;\;\;\;\;\;\;\;\;\;\;\;\;\;\;\;\;\;\;\;\;\;\;\;\;\;
\;\;\;\;\;\;\;\;\;\;\;\;\;\;\;\;\;\;\;\;\;\;
(\; \stackrel{(\ref{cubic y_2})}{\Longrightarrow}\;\;
y^3+2(4\beta^3+9i\xi)=0)
\eeqa
{\bf \underline{Proof of (\ref{basic relation}) - (\ref{cons 4})}} 

(\ref{real G_tau}) implies (\ref{basic relation}). 
This makes the first two terms in (\ref{imag G_k}) cancel
\beqa
0=\frac{a_k}{2p}Re\, W^k \Big( -2\frac{1}{y-2\beta \C}
+(2\G -1)\frac{-2\beta\C}{\cB (y-2\beta\C)}\Big)
-a_k\frac{Im\, \big[ W^k (-ip\cB +\bar{W})\big]}{|ip\cB+W|^2}
\eeqa
which proves (\ref{cons 1}).
With (\ref{k derivatives real}) and 
(\ref{basic relation}) one gets that in (\ref{real G_k})
the first two terms cancel so that  one gets the following relation
implying (\ref{cons 2}) using $Re \, W =0$
\beqa
\frac{s_k}{a_k}=\frac{Re\, \big[ W^k (-ip\cB +\bar{W})\big]}{|ip\cB+W|^2}
\eeqa
Rewriting (\ref{imag G_tau}) with (\ref{basic relation}) and
multiplying by $-2\tau_2 y^2\C$ gives (\ref{cons 4}).$\; \Box$

By (\ref{cons 4}) one gets $y$ as function of $\beta$ (i.e. of $\tau_2$);
similarly for $\C$, $\cB$ and $\G$ 
(let $\zeta:=9i\xi$)
\beqa
y&=&\big[ -2(4\beta^3+\zeta)\big]^{1/3}\\
\C&=&-\frac{2\beta^2}{y^2}
=-\frac{2\beta^2}{\big[ -2(4\beta^3+\zeta)\big]^{2/3}}\\
\label{cB in beta}
\cB&=&\beta\C-y = \big[ -2(4\beta^3+\zeta)\big]^{1/3}
\Big(\frac{\beta^3}{(4\beta^3+\zeta)}-1\Big)\\
\label{G in beta}
\G&=&\frac{y}{2\beta\C}=\frac{4\beta^3+\zeta}{2\beta^3}
\eeqa
So there remain three equations for the unknowns $t_1, t_2, \tau_2$.
When one has obtained $t_1, t_2, \tau_2$ one will get $\tau_1$
from the definition (\ref{tilde C}) of $\C$.

\newpage
 
One remains with the three equations 
(for (\ref{tau2 sector}) note $\sum \frac{s_k}{a_k}
=\frac{Im\, W}{p\cB+Im\, W}=1-\G$)
\beqa
\label{k sector}
\frac{s_k}{a_k}&=&\frac{Im \, W^{(k)}}{p\cB+Im\, W}\\
\label{tau2 sector}
\S:=\sum \frac{s_k}{a_k}&=&-\frac{2\beta^3+\zeta}{2\beta^3}
\eeqa
Now let us decouple the $t_i$ sector from the $\tau_2$ sector.
(\ref{k sector}) gives first
\beqa
\label{prop relation}
-\sqrt{\frac{t_2}{t_1}}=\frac{s_2}{s_1}&=&
\frac{a_2}{a_1}\; 
\frac{\pm_2 \, C_2 \, e^{-a_2 t_2}}{\pm_1 \, C_1 \, e^{-a_1 t_1}}\\
\label{S relation}
\S&=&\frac{Im \, W}{p\cB+Im\, W}\;\; \;\;\;\;\;\;\;\;\;\; \;\;\;\; \;\;\;\; 
\big( \mbox{with} \;\;\;
\cB=\sqrt[3]{-2\zeta\frac{\S-1}{\S+1}}\;\;\frac{\frac{1}{2}-\S}{\S-1}\big)\;\;
\eeqa
For (\ref{S relation}) note (\ref{cB in beta}) and
that (\ref{tau2 sector}) gives
$\beta^3=-\frac{1}{2}\frac{\zeta}{1+\S}$; the latter relation will
give $\tau_2$ once we have solved the two 
equations (\ref{prop relation}) and (\ref{S relation}) for the $t_i$.

These equation can also be written as 
(note $\S=\sum \frac{s_k}{a_k}$ and $Im\, W = \sum Im \, W^{(k)}$)
\beqa
\label{final first}
\frac{s_2/a_2}{s_1/a_1}&=&\frac{Im \, W^{(2)}}{Im \, W^{(1)}}\\
\label{final second}
Im \, W &=& p \sqrt[3]{-2\zeta}\;\; \frac{\S(\S-\frac{1}{2})}{\S^2-1}
\eeqa
Obviously these equations could possibly have the trivial solution that 
the common proportionality factor in (\ref{final first}) is $-1$
(if that value can be realised); this
then means however that $\S=0=Im\, W$ and $\G=1$, so that $y-2\beta\C=0$, 
according to (\ref{G in beta}),
contrary to hypothesis (it would imply $c=0=d$). 

Let us recall finally also the explicit expressions for the 
independent variables $t_k$
\beqa
s_k= (-1)^{k+1}\;\frac{3}{2}\; \frac{t_k^{1/2}}{t_2^{3/2}-t_1^{3/2}} \;\;\; 
 ,  \;\;\;
t_k = - \frac{3}{2}\; \frac{s_k^2}{s_1^3+s_2^3}\;\;\; ; \;\;\;
Im \; W^{(k)}=\pm_k \, C_k \, e^{-a_k t_k}
\eeqa
The solution to (\ref{final first}), (\ref{final second}) is
(we take $\pm_k=+$, it can be absorbed in $C_k$ anyway)
\beqa
a_k t_k = \frac{a_k^3}{a_2^3-a_1^3}\, L \;\;\; 
\Longrightarrow \;\;\; \frac{s_k}{a_k} = (-1)^{k+1} \frac{3}{2L}\;\; , \;\; 
Im W^{(k)} = (-1)^{k+1} \frac{C_1^{\frac{a_2^3}{a_2^3-a_1^3}}}
{(-C_2)^{\frac{a_1^3}{a_2^3-a_1^3}}}
\eeqa
(where $L:=\ln ( - C_2/C_1)$)
such that (note the independence of $p$) one has only the mentioned 
trivial solution $\S=0=Im\, W$. Thus there are no SUSY solutions.

Note that the conclusion would not change if one includes more generally
a proper flux $E=e+e' \tau$ (not just to mimic formally $W^{(T)}$): one gets
then for $\beta {\cal \C}:=\beta C + E + W^{(T)}$ that 
$(\beta {\cal \C})_{\tau}=\beta_{\tau}C_E + \beta \frac{c/2}{a-3b}$ with
$C_E=C+\frac{g}{d}e'$ leaving (\ref{cons 4}) intact.

\newpage

\subsection{The case III: $g+\frac{1}{3}h=0$}

For the flux components (again sticking to the case $c=3d$)
\beqa
(e_R^a \, | \, m_R^a)= (0, g , h \, | \, 0 , -a , -b)
\;\;\; , \;\;\;
(e_{NS}^a \, | \, m_{NS}^a) = (0 , 0 , 0 \, | \, 0 , c , d)
\eeqa
one gets the flux superpotential $W_{flux}=gF_1+hF_2+Az_1+Bz_2$ or
\beqa
W_{flux}&=&g\Big( -\frac{1}{2}Z^2+\frac{3}{2}Z+\frac{17}{4}\Big) 
+\frac{h}{3}\Big(-\frac{1}{2}(Z^2-z_2^2) + \frac{3}{2}(Z-z_2)+\frac{18}{4}\Big)
+\frac{A}{3}Z-pz_2\nonumber
\eeqa
One gets as elimination equations $\frac{1}{3}D_{z_1}W=0$ and $D_{z_2}W=0$
\beqa
\label{Case 2 z_1 elim}
g(-Z+\frac{3}{2})+\frac{h}{3}(-Z+\frac{3}{2})+\frac{A}{3}&=&
\frac{3}{2i}\frac{Y^2}{Y^3-y^3+9i\xi}\; W_{flux}\\
g(-Z+\frac{3}{2})+\frac{h}{3}(-Z+z_2)+B&=&
\frac{3}{2i}\frac{Y^2-y^2}{Y^3-y^3+9i\xi}\; W_{flux}
\eeqa
and the compatibility equation reads
\beqa
\Big[ \frac{h}{3}\big(\frac{3}{2}-z_2\big) +p\Big]Y^2 
= \Big[ \big(g+\frac{h}{3}\big)\big(-Z+\frac{3}{2}\big)+\frac{A}{3}\Big]y^2
\eeqa
Now let us introduce again a linear binding between $g$ and $h$. 
Whereas previously we had set
$h=0$ we put now $g+\frac{h}{3}=0$. This gives
\beqa
\label{z_2 in Y}
\Big[ \frac{h}{3}\big(\frac{3}{2}-z_2\big) +p\Big]Y^2 = \frac{A}{3}y^2\;\;\;
\stackrel{y=Y^2/\beta}{\Longrightarrow}\;\;\;
g\big(z_2-\frac{3}{2}\big)+p=\frac{A}{3}\big(\frac{Y}{\beta}\big)^2
\eeqa
(taking the imaginary part gives the indicated relation for $y$).
One gets then for $W_{flux}\, $ 
\beqa
W_{flux}
&=&g\Big( -\frac{1}{2}z_2^2 + (\frac{3}{2}-\frac{p}{g})z_2-\frac{1}{4}\Big)
+\frac{A}{3}Z
%\nonumber\\
%&=&g\Big( \frac{1}{32}-\frac{3p}{2g}+\frac{p^2}{2g^2}\Big)
%-\frac{1}{2g}\Big[ g\big(z_2-\frac{3}{2}\big)+p \Big]^2
%+\frac{A}{3}Z
\eeqa
%the latter rewritten with the help of (\ref{z_2 in Y}).
Equating with (\ref{Case 2 z_1 elim}) gives
\beqa
\label{equating the W_{flux}'es}
 \frac{A}{3}\; \frac{2i}{3}\; \frac{Y^3-\beta^{-3}Y^6+9i\xi}{Y^2}=
g\Big( \frac{1}{32}-\frac{3p}{2g}+\frac{p^2}{2g^2}\Big)
-\frac{1}{2g}\Big[ g\big(z_2-\frac{3}{2}\big)+p \Big]^2
+\frac{A}{3}Z\;\;\;
\eeqa
One gets thereby as effective 'superpotential' 
\beqa
\label{Weff case 2}
W_{eff}=\frac{A}{3}\; \frac{2i}{3}\; \frac{Y^3-\beta^{-3}Y^6+9i\xi}{Y^2}
\eeqa
where $Y$ is determined by (gotten from taking $A_2 \Re - A_1 \Im$
of (\ref{equating the W_{flux}'es}) to eliminate $X$)
\beqa
\label{sextic}
\beta^{-3}Y^6+2Y^3+CY^2-4\cdot 9i\xi=0
\eeqa
One can not tune this to become a quadric in $Y^3$ 
as the demand for vanishing of 
\beqa
C=-\frac{27}{16}\frac{\beta}{A_1^2+A_2^2}(g^2-48pg+16p^2)
\eeqa
leads to the condition
$g=4(6\pm \sqrt{35})p$, which can not be solved in integral fluxes.
\newpage
However including again the fluxes $m_R^0=-e, \, m_{NS}^0=e'$ and 
the ensuing term $E=e+e'\tau$ in $W_{flux}$ leads effectively to 
the following coefficient in the sextic
\beqa
\label{first C(E) expression}
\tilde{C}&=&-\frac{27}{16}\frac{\beta}{A_1^2+A_2^2}
\Bigg[\Big(g^2-48pg+16p^2\Big) +\frac{32g}{A_2}
\big( A_2 E_1 - A_1 E_2\big) \Bigg]\\
&=&-\frac{27}{16}\frac{\beta}{A_1^2+A_2^2}
\Big(g^2-48p\Big[ 1-\frac{2(e-\frac{a}{c}e')}{3p} \Big]g+16p^2\Big)
\eeqa
This is easily solved in special cases, such as 
$e-\frac{a}{c}e'=2p$ giving $g=-4p$ leading to the sextic being just
quadratic in $Y^3$. However as the $E$-sector is used by us to include
the $W^{(\T)}$-dependence formally, giving the $z_i(\tau, \T_i)$, 
any special tuning of it can not be kept under the further procedure, 
such that one will have still to face the more general sextic. 
When the $E$-sector is used for this purpose (to just mimic the $W^{(\T)}$) 
one has to use just the expression (\ref{first C(E) expression}). 
So one has now to work with the sextic with $C$ replaced by $\tilde{C}$.

Now $\beta = \frac{d}{g}\tau_2,$ gives $\beta_{\tau}=\frac{d}{2ig}$,
$\beta_{\T_k}=0$ and
(\ref{first C(E) expression}) gives
%($E:=W^{(\T)}=\sum W^k=\sum C_k e^{-a_k\T_k}$)
%\beqa
%\C_{\tau}&=&-\frac{27}{16}\frac{\beta}{A_1^2+A_2^2}
%\Big\{ (\frac{1}{2i\tau_2}-\frac{c}{A})
%\Big[\Big(g^2-24pg+16p^2\Big) +32g
%\big( E_1 - \frac{A_1}{A_2} E_2\big) \Big]
%-\frac{c}{A}\, 32g\, \frac{i}{2}\frac{A\bar{A}}{A_2^2}E_2 \Big\}\nonumber\\
%&=&\frac{9}{A_1^2+A_2^2}
%\Big\{ (\frac{i}{2\tau_2}+\frac{c\bar{A}}{A\bar{A}})
%A_2\Big[Q 
%-2\big( E_1 - A_1 \frac{E_2}{A_2}\big) \Big]
%+ic\bar{A}\frac{E_2}{A_2} \Big\}\nonumber\\
%\C_{\T_k}&=&-\frac{27}{16}\frac{\beta}{A_1^2+A_2^2}\; \frac{32g}{A_2}\,
%\frac{i}{2}\bar{A}\, (-a_k)\, W^k = i\frac{9}{2A}\, a_k W^k
%\nonumber
%\eeqa
%(with $Q:=\frac{g}{16}-\frac{3}{2}p+\frac{p^2}{g}$).
$\C_{\T_k}= i\frac{9}{2A}\, a_k W^k$ and a more complicated expression
for $\C_{\tau}$.
From (\ref{sextic}) with $C\ra \tilde{C}$ one finds 
$Y_l\big( \beta^{-3}Y^4+Y+\frac{\tilde{C}}{3}\big)
=\frac{1}{2}\big(\beta^{-4}\beta_l Y^4 -\frac{\tilde{C}_l}{3}\big) Y$
(here $l$ is $\tau$ or $\T_i$).
With the notation
$\cB:=-\beta^{-3}Y^4+2Y+\frac{\tilde{C}}{3}$
one gets for the effective Kahler potential
(up to additive constants) and superpotential
\beqa
K_{eff}(\tau)&=&-\ln\Big( Y^3 - y^3 + 9 i\xi\Big)=-2\ln Y - \ln \cB\\
\label{Weff in case 3 to refer}
W_{eff}(\tau)&=&=i\cA\; \cB\;\;\;\;\;\;\;\;\; (\cA:=\frac{A}{6})
\eeqa
One gets as effective $G$-function and SUSY conditions (note the marked 
difference in this case that now
$A\in {\bf C}$ is $\tau$-dependent 
whereas earlier $p\in {\bf R}$ was a constant)
\beqa
G&=&-\ln(\tau-\bar{\tau})-2\ln Y - \ln \cB 
+\ln\big|i\cA\; \cB +W^{(\T)}\big|^2 +K(\T_k)\\
G_k&=&-2\frac{y_k}{y}+(2\G-1)\frac{\cB_k}{\cB}-a_kW^k\; \G'+s_k=0\\
G_{\tau}&=&-2\frac{y_{\tau}}{y}
+(2\G-1)\frac{\cB_{\tau}}{\cB}+i\frac{c}{6}\cB\; 
\Gamma'+\frac{i}{2\tau_2}=0
\eeqa
where we use the abbreviations
\beqa
\Gamma=\cB\, 
\frac{\cA_1\big(\Im \, W^{(T)}+\cA_1 \cB\big)
-\cA_2\big(\Re\, W^{(T)}-\cA_2 \cB\big)}
{|i\cA \cB+ W^{(T)}|^2},
\Gamma'=\frac{-i\big(\Im \, W^{(T)}+\cA_1 \cB\big)
+\big(\Re\, W^{(T)}-\cA_2 \cB\big)}
{|i\cA \cB+ W^{(T)}|^2}
\nonumber
\eeqa
One can now start to proceed as in the previous examples. We will
not write the ensuing equations here, as they become not especially
illuminating. However, by now the general scheme should be rather clear.
Proceeding in the manner described, the interested reader should be able
to use this method to work out further cases. Note that each case will
be of a somewhat different nature (cf. the remark here after 
(\ref{Weff in case 3 to refer})), and furthermore although 
for more general flux combinations
the existence of SUSY vacua becomes more likely, 
on the other hand the explicit 
determination of $W_{eff}$ also becomes more involved.

\newpage

\section{Discussion and Outlook}

\resetcounter

One of the most often studied scenarios of moduli stabilization
is the KKLT set-up. This leads to supersymmetric AdS vacua with all
moduli stabilized. Eventually this is uplifted to a SUSY-breaking
dS vacuum. As the uplift process is not easily treated in a controlled
framework other approaches to get moduli stabilization and dS vacua
are of course of the highest interest [\ref{CK1}], [\ref{CK2}].

In this paper we stick just to the moduli stabilization aspect
of the KKLT scenario. The common procedure employed in the KKLT
framework is to use first [\ref{GKP}] to fix the complex structure
moduli $z_i$ and the dilaton $\tau$ from a flux superpotential $W_{flux}$
and then, afterwards, 
to use non-perturbative effects to stabilize the K\"ahler moduli $\T_k$.
The underlying assumptions of the decoupling in this two-step procedure
have been questioned [\ref{Nilles}] (cf.~also [\ref{CKL}]).
This led to a procedure where only the $z_i$ are integrated out
what leads to an effective superpotential $W_{eff}(\tau)$
(in [\ref{Nilles}] conditions for a valid uplift were also considered).
As this $W_{eff}(\tau)$ is a non-holomorphic quantity, as already pointed
out in [\ref{de Alwis}], the whole procedure actually has to be rephrased
with the aid of the $G$-function of supergravity. When the integrating 
out-process is incorporated appropriately one actually gets not only a
$W_{eff}(\tau)$ but a $W_{eff}(\tau, \T_k)$; it then contributes 
via the $\ln| W_{eff}(\tau, \T_k)+W^{(\T)}(\T_k)|^2$ term in the effective 
$G$-function.

This program, which was described in [\ref{Nilles}] and [\ref{de Alwis}]
in a generic effective supergravity framework, is carried out here
for one of the most intensely studied Calabi-Yau related models,
the orientifold model of ${\bf P}_{11169}[18]$, which was also
studied in the original KKLT framework [\ref{DDF}]. Even more,
this geometry was also the background to study further questions,
such as inclusion of $\al'^3$-corrections [\ref{CoQu}], [\ref{BaBe}] 
or inflation [\ref{8authors}].

The new approach is investigated here for this geometry in various
flux combinations. More precisely, the determination of $W_{eff}$ 
is illustrated mainly for the case
\beqa
(e_R^a \, | \, m_R^a)= (0, g , 0 \, | \, 0 , -a , -b)
\;\;\; , \;\;\;
(e_{NS}^a \, | \, m_{NS}^a) = (0 , 0 , 0 \, | \, 0 , c , d)
\eeqa
One gets for $g\neq 0$ as effective superpotential (the case II above)
\beqa
W_{eff}(\tau, \T_k)&=&i\frac{a-3b}{3}\Big( \frac{d}{g}\tau_2 
\frac{a+c\tau_1 + \frac{9}{2}g
+\frac{3g}{d\tau_2}\mbox{Im}W^{(\T)}(\T_i)}{a-3b}-y\Big)
\eeqa
Here $y$ is given as function of $\tau$ and $\T_k$ 
by the Cardano formula for the cubic
\beqa
0&=&y^3-\Big[ 3\frac{d}{g}\tau_2 \frac{a+c\tau_1 + \frac{9}{2}g
+\frac{3g}{d\tau_2}\mbox{Im}W^{(\T)}(\T_i)}{a-3b}\Big] y^2 
+ 2\Big( (\frac{d}{g}\tau_2 )^3+9i\xi\Big)\;\;\;\;
\eeqa
Here $W^{(\T)}$ denotes $\sum_{k=1}^2 C_k e^{-a_k\T_k}$. 

Above we did assume that $g\neq 0$. In the case
$g=0$ one gets (case I above)
\beqa
W_{eff}(\tau, \T_k)&=&i(d\tau+\frac{a}{3})\Big( Y + 3\, 
\frac{(a+c\tau_1)\Im \, W^{(\T)}-c\tau_2\Re\, W^{(\T)}}{|a+c\tau|^2}\Big)
\eeqa
Again, here $Y$ is given as function of $\tau$ and $\T_k$ 
by the Cardano formula for the cubic
\beqa
0&=&Y^3+9\, 
\Big[ \frac{(a+c\tau_1)\Im \, W^{(\T)}-c\tau_2\Re\, W^{(\T)}}{|a+c\tau|^2}
\Big] Y^2
-18i\xi
\eeqa
In these cases one can resolve analytically
the remaining supersymmetry conditions
on $\tau$ and the $\T_k$. As the case $g=0$ is
somewhat degenerate ($N_{flux}=0$) we give the 
treatment fully only for our main case $g\neq 0$.
There one finds that no supersymmetric solutions to the ensuing
supersymmetry conditions for $\tau$ and the $\T_k$ exist. 

There is a third, more complicated case with a non-zero $e_R^2$ turned on
(the case III above),
where we also give the effective superpotential and pave the way for
the treatment of the supersymmetry conditions. 
As in the present paper our aim was just to elucidate the concrete
improved KKLT procedure in the spirit of [\ref{Nilles}] and [\ref{de Alwis}],
we are content here to illustrate this new set-up with a number of
non-trivial examples where actually all the steps can be carried out
in closed analytical form (this program is surely completely worked out
at least for the main case $g\neq 0$, i.e. case II).

The worked out examples should make it sufficiently clear
how the preceding considerations could be generalised further.
One next step would be to 'solve' the ${\bf P}_{11169}[18]$
model in the sense of allowing (more) general flux combinations.
It would be especially interesting to turn on the flux $e_R^0$
which leads to a contribution $e_R^0\, F_0$ in $W_{flux}$
which would be even of cubic degree in the $z_i$ and so will lead
to a more complicated dependence in $W_{eff}$. As the treatment [\ref{DDF}]
along the lines of the original KKLT set-up contains this flux
this would open up the possibility for a direct comparison. Of course
alternatively one can study the KKLT procedure also in one of the simpler 
examples by demanding $D_{\tau}W_{flux}=0$ in addition to 
$D_{z_i}W_{flux}=0$. We indicated this in 
(\ref{KKLT procedure integrating out}) and a similar analysis
is easily given for $g\neq 0$. Another interesting avenue to explore
would be to combine the appropriate modification of the KKLT procedure
outlined here with further questions such as the inclusion of
$\al'^3$-corrections [\ref{CoQu}], [\ref{BaBe}] 
or inflation [\ref{8authors}]. Furthermore one could carry out the program 
described here for other examples of low parameter Calabi-Yau 
(orientifold) models. Finally one could, following [\ref{Nilles}], 
work out the appropriate criterion for a successful
uplift to dS space in connection with the question of stability, i.e.
one could study the nature of the supersymmetric points with respect to
the stronger dS stability criterion. Touching thereby the realm of 
non-supersymmetric solutions one could study numerically their existence 
already for the potential arising here (without uplift).

We would like to thank S.~Theisen for participation 
at some stages of this work.

\newpage

\appendix

%\section{Appendix: Some solutions for $\tau$ integrated out}

%We recall the solutions of [\ref{DDF}] for the case of having
%integrated out the $z_i$ {\em and} $\tau$ and working just with $W_0$.

%\[ \begin{array}{cc|c||cc|cc|c}
%a_1 & a_2 & W_0 & t_1 & t_2 & \tau_1 & \tau_2 & Vol       \\
%\hline\hline
%1/4 & 1/30 & 10^{-30} & 9.83 & 2.76 & 48.3 & 348  & 484   \\
%1/4 & 1/30 & 10^{-5}  & 4.61 & 1.14 & 10.6 & 65.7 & 39.1  \\
%1/4 & 1/12 & 10^{-30} & 9.73 & 1.16 & 47.4 & 139  & 103   \\
%1/4 & 1/12 & 10^{-5}  & 4.40 & 0.468& 9.64 & 25.9 & 8.01  \\
%\end{array} \]

\section{SUSY conditions in case I}

Let us consider the real and imaginary parts of $G_k=0$ and $G_{\tau}=0$
(with $W:=W^{(\T)}$)
\beqa
\label{real G_k case I}
-2\frac{Re\, Y_k}{Y}+(2\G-1)\frac{Re\, \cB_k}{\cB}
-a_k\frac{Re\, 
\big[ W^k (-i\bar{\cA}\cB +\bar{W})\big]}{|i\cA\cB+W|^2}&=&-s_k\;\;\;\;\\
\label{imag G_k case I}
-2\frac{Im\, Y_k}{Y}+(2\G-1)\frac{Im\, \cB_k}{\cB}
-a_k\frac{Im\, \big[ W^k (-i\bar{\cA}\cB +\bar{W})\big]}{|i\cA\cB+W|^2}
&=&0\\
\label{real G_tau case I}
-2\frac{Re \, Y_{\tau}}{Y}+(2\G-1)\frac{Re\, \cB_{\tau}}{\cB}
+\frac{c}{3}\cB\frac{\cA_1\cB+\Im\, W}
{\big(\cA_1\cB+\Im\, W\big)^2+\big(-\cA_2\cB+\Re\, W\big)^2}&=&0\\
\label{imag G_tau case I}
-2\frac{Im \, Y_{\tau}}{Y}+(2\G-1)\frac{Im\, \cB_{\tau}}{\cB}
+\frac{c}{3}\cB\frac{ -A_2\cB+\Re\, W}
{\big(\cA_1\cB+\Im\, W\big)^2+\big(-\cA_2\cB+\Re\, W\big)^2}
&=&-\frac{1}{2\tau_2}\;\;\;\;\;\;\;
\eeqa

Note first that the middle two real equations $\Im \, G_k=0$ 
and $\Re \, G_{\tau}=0$,
which have a zero on the right hand side, have a major part in common
(again $W$ denotes $W^{(\T)}$)
\beqa
\label{imag G_k case I second form}
\Im \C_k \, \Big[ \frac{2}{Y+2\C}+(2\Gamma -1) 
\frac{2\C}{Y+2\C}\frac{1}{Y+\C} \Big]&=&a_k
\frac{\Im\big( W^k(-i\bar{\cA}\cB+\bar{W})\big)}{|i\cA \cB +W|^2}\\
\label{real G_tau case I second form}
\Re \C_{\tau}\, \Big[ \frac{2}{Y+2\C}+(2\Gamma -1) 
\frac{2\C}{Y+2\C}\frac{1}{Y+\C} \Big]&=&-\frac{c}{3}\cB
\frac{\cA_1\cB+\Im W}{|i\cA \cB +W|^2}
\eeqa
Using (\ref{real G_tau case I second form}) 
in (\ref{imag G_k case I second form}) gives
\beqa
-\frac{c}{3}\cB\; \frac{\Im \C_k}{\Re \C_{\tau}}\;\big(\cA_1 \cB +\Im W\big)
=a_k \Big( -\Re W^k (\cA_1 \cB + \Im W)+\Im W^k (-\cA_2 \cB +\Re W) \Big)\;\;\;
\eeqa
Dividing by $a_k$ and summing over $k$ one arrives at
\beqa
\frac{-c\cB}{\Re \C_{\tau}}\, \frac{\cA_1\Re W + \cA_2 \Im W}{2A\bar{A}}\,
\big(\cA_1 \cB +\Im W\big)=-\cB \big( \cA_1\Re W + \cA_2 \Im W\big)
\eeqa
This implies $\cA_1\Re W + \cA_2 \Im W=0$ 
for otherwise one would get
$\cB=2\frac{\cA_2 \Re W - \cA_1 \Im W}{\cA_1^2+\cA_2^2}$ from
\beqa
\big(\cA_1 \cB +\Im W\big) (\cA_1^2+\cA_2^2)=2\cA_1\cA_2\Re W 
- (\cA_1^2-\cA_2^2)\Im W
\eeqa
But one has that
$\cB=Y-\frac{\cA_2 \Re W - \cA_1 \Im W}{\cA_1^2+\cA_2^2}$; this would then
give $Y=3\frac{\cA_2 \Re W - \cA_1 \Im W}{\cA_1^2+\cA_2^2}=-3\C$
which is excluded from (\ref{cubic equation case I})
as $\xi\neq 0$. So the remaining possibility is 
$\Im W = -\frac{\cA_1}{\cA_2}\Re W$ and so $\C=-\frac{\Re W}{\cA_2}$.
Applying now (\ref{real G_tau case I second form}) in (\ref{imag G_tau case I})
would give finally $0=-\frac{1}{2\tau_2}\frac{3}{c\cB}
\frac{\cA_1^2+\cA_2^2}{\cA_2}(\cA_2 \cB + \Re W)\frac{\cA_1}{\cA_2}\Re W$,
so $\cA_1=0$ or $\Re W=0$ (and so in any case $\Im W=0$); however
above we had assumed that $\Re \C_{\tau}\neq 0$ and so this can not be argued.

Rather one has from 
$\Im W = -\frac{\cA_1}{\cA_2}\Re W$ and $\C=-\frac{\Re W}{\cA_2}$
(still assuming $\Re \C_{\tau}\neq 0$) 
that (\ref{imag G_tau case I}) gives finally
$4\cA_2^2\cB (\Re W)^2=(\cB+2\cA_2 \Re W)
\big( (\cA_2 \cB)^2-(\Re W)^2\big)$ or
\beqa
Y^3+(3\C-2\cA_2^2\C)Y^2+2(1-4\cA_2^2)\C^2 Y + 4\cA_2^2 \cdot \C^3=0
\eeqa
Combining with (\ref{cubic equation case I}) one gets eliminations
with which $G_k=0$ can be treated further.

\section*{References}
\begin{enumerate}

\item
\label{GKP}
S.B. Giddings, S. Kachru and J. Polchinski,
{\em Hierarchies from Fluxes in String Compactifications},
hep-th/0105097, Phys. Rev. {\bf D66} (2002) 106006.

\item
\label{fluxes}
S.~Gukov, C.~Vafa and E.~Witten,
{\em CFT's From Calabi-Yau Four-folds}, hep-th/9906070, 
Nucl.Phys. {\bf B584} (2000) 69, Erratum-ibid. B608 (2001) 477;\\
S.~Gukov,
{\em Solitons, Superpotentials and Calibrations},
hep-th/9911011, Nucl.Phys. {\bf B574} (2000) 169;\\
T.~R.~Taylor and C.~Vafa,
{\em RR flux on Calabi-Yau and partial supersymmetry breaking},
Phys.\ Lett.\ B {\bf 474}, 130 (2000),
hep-th/9912152;\\
P.~Mayr,
{\em On supersymmetry breaking in string theory and its
realization in brane worlds}, Nucl.\ Phys.\ B {\bf 593}, 99
(2001), hep-th/0003198;\\
G.~Curio, A.~Klemm, D.~L\"ust and S.~Theisen,
{\em On the vacuum structure of type II string compactifications
on  Calabi-Yau spaces with H-fluxes}, Nucl.\ Phys.\ B {\bf 609}, 3
(2001), hep-th/0012213.

\item
\label{KKLT}
S. Kachru, R. Kallosh, A. Linde and S.P. Trivedi,
{\em De Sitter Vacua in String Theory}, hep-th/0301240, 
Phys. Rev. {\bf D68} (2003) 046005.

\item
\label{DRSW}
H.P. Nilles, {\em Gaugino Condensation and SUSY Breakdown}, 
hep-th/0402022;\\
S. Ferrara, L. Girardello and H.P. Nilles,
{\em Breakdown Of Local Supersymmetry Through Gauge Fermion
Condensates},
Phys. Lett. {\bf B125} (1983) 457;\\
M. Dine, R. Rohm, N. Seiberg and E. Witten,
{\em Gluino Condensation in Superstring Models},
Phys. Lett. {\bf B156} (1985) 55;\\
J.P. Derendinger, L. E. Ibanez and H. P. Nilles,
{\em On The Low-Energy D = 4, N=1 Supergravity Theory Extracted
From The D = 10, N=1 Superstring}, Phys. Lett. {\bf B155} (1985)
65;\\
J.P. Derendinger, L.E. Ibanez and H. P. Nilles,
{\em On The Low-Energy Limit Of Superstring Theories},
Nucl. Phys. {\bf B267} (1986) 365;\\
A. Font, L.E. Ibanez, D. L\"ust and F. Quevedo,
{\em Supersymmetry Breaking From Duality Invariant Gaugino
Condensation},
Phys. Lett. {\bf B245} (1990) 401;\\
S. Ferrara, N. Magnoli, T.R. Taylor and G. Veneziano,
{\em Duality And Supersymmetry Breaking In String Theory},
Phys. Lett. {\bf B245} (1990) 409;\\
H.P. Nilles and M. Olechowski,
{\em Gaugino Condensation And Duality Invariance}, Phys. Lett.
{\bf B248} (1990) 268.

\item
\label{W}
E. Witten, 
{\em Non-Perturbative Superpotentials In String Theory},
hep-th/9604030, Nucl.Phys. {\bf B474} (1996) 343.

\item
\label{Kallosh}
L. Goerlich, S. Kachru, P.K. Tripathy and S.P. Trivedi,
{\em Gaugino Condensation and Nonperturbative Superpotentials 
in Flux Compactifications},
hep-th/0407130, JHEP {\bf 0412} (2004) 074;\\
R. Kallosh and D. Sorokin, 
{\em Dirac Action on M5 and M2 Branes with Bulk Fluxes},
hep-th/0501081, JHEP {\bf 0505} (2005) 005;\\
P.K. Tripathy and S.P. Trivedi, 
{\em D3 Brane Action and Fermion Zero Modes in Presence of Background Flux},
hep-th/0503072, JHEP {\bf 0506} (2005) 066;\\
E. Bergshoeff, R. Kallosh, A.-K. Kashani-Poor, D. Sorokin and
A. Tomasiello,
{\em An index for the Dirac operator on D3 branes with background fluxes},
hep-th/0507069, JHEP {\bf 0510} (2005) 102;\\
R. Kallosh, A.-K. Kashani-Poor and A. Tomasiello,
{\em Counting fermionic zero modes on M5 with fluxes},
hep-th/0503138, JHEP {\bf 0506} (2005) 069;\\
D. Lust, S. Reffert, W. Schulgin and P. Tripathy,
{\em Fermion Zero Modes in the Presence of Fluxes 
and a Non-perturbative Superpotential},
hep-th/0509082.

\item
\label{Nilles}
K. Choi, A. Falkowski, H. P. Nilles, M. Olechowski and S. Pokorski,
{\em Stability of Flux Compactifications and the Pattern of
Supersymmetry Breaking}, hep-th/0411066, JHEP {\bf 0411} (2004) 076.

\item
\label{de Alwis}
S.P. de Alwis,
{\em Effective Potentials for Light Moduli},
Phys.Lett. {\bf B626} (2005) 223, hep-th/0506266.

\item
\label{DDF}
Frederik Denef, Michael R. Douglas, Bogdan Florea,
{\em Building a Better Racetrack},
hep-th/0404257, JHEP {\bf 0406} (2004) 034.

\item
\label{8authors}
J.J. Blanco-Pillado, C.P. Burgess, J.M. Cline, C. Escoda, M. Gomez-Reino, 
R. Kallosh, A. Linde and F. Quevedo, 
{\em Inflating in a Better Racetrack},
hep-th/0603129.

\item
\label{CoQu}
J.P. Conlon and F. Quevedo,
{\em Gaugino and Scalar Masses in the Landscape},
hep-th/0605141.

\item
\label{BaBe}
V. Balasubramanian, P. Berglund, J.P. Conlon and F. Quevedo,
{\em Systematics of Moduli Stabilisation in Calabi-Yau Flux Compactifications},
hep-th/0502058, JHEP {\bf 0503} (2005) 007.

\item
\label{G}
A. Grassi, 
{\em Divisors on elliptic Calabi-Yau 4-folds 
and the superpotential in F-theory, I}, alg-geom/9704008.

\item
\label{KV}
S. Katz and C. Vafa,
{\em Geometric Engineering of N=1 Quantum Field Theories},
hep-th/9611090, Nucl.Phys. {\bf B497} (1997) 196.

\item
\label{FMW}
R. Friedman, J. Morgan and E. Witten,
{\em Vector Bundles And F Theory},
hep-th/9701162, Commun.Math.Phys. {\bf 187} (1997) 679.

\item
\label{CFKM}
P. Candelas, A. Font, S. Katz and D.R. Morrison,
{\em Mirror Symmetry for Two Parameter Models -- II},
hep-th/9403187, Nucl.Phys. {\bf B429} (1994) 626.

\item
\label{WessB}
J. Wess and J. Bagger, {\em "Supersymmetry and Supergravity"},
Second Edition, Revised and Expanded (1992), Princeton Series in Physics,
Princeton University Press, Princeton, New Jersey.

\item
\label{CK1}
M. Becker, G. Curio and A. Krause,
{\em De Sitter Vacua from Heterotic M-Theory},
hep-th/0403027, Nucl.Phys. {\bf B693} (2004) 223.

\item
\label{CK2}
G. Curio and A. Krause,
{\em Moduli Stabilization in New Heterotic de Sitter Vacua},
to appear.

\item
\label{CKL}
G. Curio, A. Krause and D. Lust,
{\em Moduli Stabilization in the Heterotic/IIB Discretuum},
hep-th/0502168, Fortsch.Phys. {\bf 54} (2006) 225.

\end{enumerate}

\end{document}